\documentclass[11pt]{article}
\usepackage[dvips]{graphicx}
\usepackage{graphics}
\usepackage{graphicx}
\usepackage{amssymb}
\usepackage{amsmath}
\usepackage{amssymb}
\usepackage{overcite}
\usepackage{epsf}
\input{epsf}
\begin{document}
\begin{titlepage}
\title{\bf Unconventional and Exotic Magnetism in Carbon-Based  Structures and 
Related Materials\thanks{International Journal of Modern Physics B (IJMPB), Volume: 27, Issue: 11 
(2013)  p.1330007 (40 pages),\qquad  DOI: 10.1142/S0217979213300077  }}  
\author{A. L. Kuzemsky 
\\
{\it Bogoliubov Laboratory of Theoretical Physics,} \\
{\it  Joint Institute for Nuclear Research,}\\
{\it 141980 Dubna, Moscow Region, Russia.}\\
{\it E-mail:kuzemsky@theor.jinr.ru} \\
{\it http://theor.jinr.ru/\symbol{126}kuzemsky}}
\date{}
\maketitle
\begin{abstract}
The detailed analysis of the problem of possible magnetic behavior of the carbon-based structures was fulfilled
to elucidate and resolve (at least partially) some unclear issues.
It was the purpose of the present paper to look somewhat more 
critically into some conjectures which have been made  and to the  peculiar and contradictory 
experimental results  in this rather indistinct and disputable field. Firstly the basic physics of magnetism was
briefly addressed. Then a few basic questions were thoroughly analyzed and critically reconsidered to elucidate 
the possible relevant mechanism (if any) which may be responsible for observed peculiarities of the "magnetic" behavior in these systems. 
The arguments supporting the existence of the intrinsic magnetism in carbon-based materials, including pure 
graphene were analyzed critically.
It was concluded that recently published works have shown clearly that the results of the previous studies,
where the "ferromagnetism" was detected in  pure graphene,  were incorrect.
Rather, graphene is strongly diamagnetic, similar to
graphite. Thus the possible traces of a quasi-magnetic behavior which some authors observed in their samples
may be attributed rather to induced magnetism due to the impurities, defects, etc.
On the basis of the present analysis the conclusion was made that the thorough and detailed 
experimental studies of these problems only may shed light on the very complicated problem 
of the magnetism of carbon-based materials. Lastly the peculiarities of the magnetic behavior of some related materials 
and the trends for future developments were mentioned.\\ 
\vspace{0.3cm}

%
%
%
%
%
\end{abstract}
\end{titlepage}
\newpage
\tableofcontents
\newpage
%
%
%
%
\section{Introduction}\label{intro}
%
%
Magnetism is a subject of great importance which has been studied 
intensely~\cite{vons71,dorf61,tyab,cra95,bert98,coey99,hand99,blund01,gigno02,gig05,boer03,kor07,rod06,sieg06,magmat08,stef08,coey09,spald11}.
Many fundamental questions were clarified and answered and many applications were elaborated.
Various magnetic materials~\cite{coey99,hand99,magmat08,spald11}, e.g. $AlNiCo,$  samarium-cobalt, neodymium-iron-boron, hard ferrites etc., were devised which
found numerous technical applications. In particular $NdFeB$ magnets are characterized by exceptionally strong magnetic 
properties and by exceptional resistance to demagnetization.
This group of magnetic substances provides the highest available magnetic energies of any material. Moreover, $NdFeB$ 
magnets allow small shapes and sizes and have multiple uses in science, engineering and industry.\\
In the last decades many new growth points in magnetism have appeared as well. The search for macroscopic magnetic 
ordering in exotic and artificial 
materials  and  devices has attracted big attention~\cite{rod06,sieg06,magmat08,stef08,coey09,spald11,stef12}, 
forming a new branch in the condensed matter physics. \\
The development of experimental techniques and solid state chemistry~\cite{rao97} over the recent years opened the possibility for
synthesis and investigations of a wide class of new substances and artificial magnetic structures with unusual 
combination of magnetic and electronic properties~\cite{sieg06,magmat08,stef08,coey09,spald11,stef12,bram03,wut07}. 
This gave a new drive to the magnetic researches due to the finding of new magnetic materials for use as 
permanent magnets, sensors and in magnetic recording devices~\cite{hei93,sco03,scom06,mil06,zab10}.\\
In particular, the carbon-based materials~\cite{chem95,carb1,fuller93,fuller11,carb2,carb3} like graphite, fullerenes and graphene were pushed into 
the first row of researches. 
Graphene is a monolayer of carbon atoms packed into a dense
honeycomb crystal structure, which can be obtained by mechanical exfoliation from graphite~\cite{geim07,katz07,geim09,rao09,geim11,nov11,wsnano11,katz12}.
Graphene has attracted a great interest in material science due to its 
novel electronic structure.  Electrons in graphene possess many fascinating properties not
seen in other materials.  
Graphene sample can be considered as an infinite molecule of carbon atoms with two-dimensional $sp^2$ network over a honeycomb lattice.  
Electrons in graphene are not governed by the Schr\"{o}dinger equation with renormalized mass but should be described 
in terms of of a relativistic theory using the Dirac equation with vanishing mass. Thus the electrons in graphene are, 
in a sense, "relativistic particles" in condensed matter. \\
The minimum unit cell of graphene contains two equivalent carbon sites A and B. Then it becomes a
semiconductor but the energy gap vanishes at two momenta $K$ and $K'$ in the Brillouin zone, thus forming a zero-gap
semiconductor. In contrast to a conventional semiconductor, the gap linearly vanishes and the energy bands form a
cone structure (Dirac cone). The electronic structure of graphene has a topological
singularity at the Dirac point where two bands cross each other, and it gives rise to anomalous behavior
in the conductivity,  dynamical transport,  and the Hall effect. \\
Thus graphene can be viewed as the two-dimensional form of pure $sp^2$ hybridized 
carbon~\cite{chak10}. In a sense it can be considered as  a giant molecule  of atomic thickness.  In line
with theoretical predictions, charge carriers in graphene behave
like massless Dirac fermions, which is a direct consequence of the linear energy dispersion relation. \\
This results in the observation of
a number of very peculiar electronic properties, from an anomalous
quantum Hall effect to the absence of localization  in this 
two-dimensional material. It also provides a bridge between condensed
matter physics and quantum electrodynamics, and opens new perspectives for carbon-based electronics~\cite{haddon12}.
Such features are very much promising for
the use of graphene for mechanical, thermal, electronic, magnetic,
and optical applications, in spite that the absence of a band-gap in graphene seem
makes it unsuitable for conventional field effect transistors. However recently
the obstacle to the use of graphene as an alternative to silicon electronics  
(the absence of an energy gap between its conduction and valence bands, which makes it difficult to achieve low power 
dissipation in the off state) has been overcame. It was reported~\cite{geim12} about fabrication of a bipolar field-effect 
transistor that exploits the low density of states in graphene and its one atomic layer thickness. The prototype 
devices are graphene heterostructures with atomically thin boron nitride or molybdenum disulfide acting as a vertical 
transport barrier. They exhibit high room 
temperature switching ratios. Thus such devices may have potential for high-frequency 
operation and large-scale integration. \\ 
It was conjectured in the last decade that in addition to its transport properties~\cite{bost07,aus07,bas08,gra1,gra2,muc10} a rich variety of magnetic
behavior may be expected in carbon-based materials and  graphene, including even a kind of intrinsic ferromagnetism. Some hypothesis were claimed  
that connected possible spin-ordering effects with the low-dimensionality and Dirac-like electron spectrum of graphene,
thus inspiring a new kind of   magnetism  without magnetic ions.  
Indeed, the understanding and control of the potential magnetic properties of carbon-based materials 
may be of fundamental relevance in applications in nano- and biosciences. However the problem is not solved yet.\\
In spite of the fact that magnetism is not usually expected in simple $sp$ oxides like $MgO$ or in carbons like graphite 
it was speculated that  basic intrinsic defects in these systems~\cite{ston10,yazy10,he12} may be magnetic in ways 
that seem to be shared by more complex oxides. 
The possible magnetic nature of these intrinsic carbon defects may suggests that it is important to understand their role 
in the recently reported  "magnetism" in some carbon-based systems.
Moreover, a "room-temperature ferromagnetism of graphene" was claimed~\cite{wang09}. However, the mechanism responsible 
for that "ferromagnetism" in carbon-based materials, which contain only $s$ and $p$ electrons in
contrast to traditional ferromagnets based on $3d$ or $4f$ electrons, is still rather unclear.\\
Thus the natural question arises:
can carbon-based materials be magnetic in principle and what is the mechanism of the appearing of the
magnetic state from the point of view of the quantum theory of magnetism? In addition, it should be emphasized 
strongly that almost all of the properties of these substances are affected by the imperfections and impurities
of the nano-structures.\\ 
In the present work,
these questions were analyzed and reconsidered to elucidate the possible relevant mechanism (if
any) which may be responsible for observed peculiarities of the "magnetic" behavior in these
systems, having in mind the quantum theory of magnetism criteria~\cite{kuz09,kuz10}. Emphasis is placed on revealing key concepts and
really measured magnetic phenomena on which such speculations rest.
%
%
%
%
%
\section{Magnetism and Magnetic Materials}
%
%
%
Before taking up the problem of the magnetic behavior of the carbon-based structures we must summarize briefly the
most relevant of the fundamental concepts of the physics of magnetism.
There are many examples of physical systems with a stable magnetic moment in the ground state~\cite{vons71,tyab,kor07,coey09}. These systems 
are the atoms, molecules and ions with an odd number of electrons, some molecules with an even number of 
electrons ($O_{2}$  and some organic compounds) and atoms (ions) with an unfilled ($3d-$, $4f-$, $5f-$) shells.
Within each shell, electrons can be specified according to their orbital angular momenta, $s-$electron having no angular momentum,
$p-$electrons having one quantum of angular momentum, $d-$electrons having  two, and   $f-$electrons having three. The
$s-$ and $p-$states tend to fill before $d-$states as the atomic number $Z$ increased. Each electron carries with it as it moves a 
half quantum of intrinsic angular momentum or spin~\cite{tomo97,sten02} with an associated magnetic moment. It have only two orientations relative
to any given direction, parallel or antiparallel.\\
Magnetic materials~\cite{hand99,sieg06,magmat08,stef08,coey09,spald11,stef12},  as a rule,   can be metals, semiconductors or insulators
which contain the ions of the transition metals or rare-earth metals with 
unfilled shells~\cite{vons71,cra95,blund01,coey09,kuz09}.Strong magnetic materials as a rule include $3d-$ ions with 
unfilled shells~\cite{kuz09}. The question of formulation of the universal criterion for \emph{ferromagnetism} is a difficult
problem, because of the existence of the huge variety of the magnetic substances and structures.\\
According to Pauli exclusion principle~\cite{pauli05},
the electrons with parallel spins tend to avoid each other spatially.
One can say that the Pauli exclusion principle lies in the foundation of the quantum theory of
magnetic phenomena. 
It is worth noting that the magnetically active electrons which form the magnetic
moment can be localized or itinerant (collectivized)~\cite{kuz09,kuz81,akuz02,kuz02,her66,tmor88,kubl,mizia}.\\
The origin of magnetism lies in the orbital and spin motions of electrons and how the electrons interact 
with one another~\cite{vons71,dorf61,tyab,cra95,blund01,coey09,kuz09}. The basic object in the magnetism of condensed matter
is the magnetic moment $\textbf{M}$. It can be imagined as a magnetic dipole. 
Magnetic moment $\textbf{M} = g \mu_{B} \textbf{J}$ is proportional to the total angular 
momentum $\textbf{J} = \textbf{L} + \textbf{S},$  where $\textbf{L}$ is orbital moment and $\textbf{S}$ is spin moment.
In nature, magnetic moments are carried by magnetic minerals the most common of which are magnetite and 
hematite~\cite{vons71,magmat08,coey09,spald11}.
The magnetic moment in practice may depend 
on the detailed environment and additional interactions such as spin-orbit, screening effects and crystal fields. 
%
%
%
%
%
%
\begin{table}
\label{tb1}
\begin{center}
Table 1. Magnetic behavior of materials.
\end{center}
\begin{center}
\begin{tabular}{|l|l|l|} \hline 
Types of magnetism & Net magnetic moment & Order/disorder \\[0.8ex] 
\hline \hline\\[-1.8ex] 
Diamagnetism&absent&no long-range magnetic order\\
\hline 
Paramagnetism&absent&no long-range magnetic order\\
\hline 
Ferromagnetism&strong&long-range magnetic order for $T < T_{c}$\\
\hline
Ferrimagnetism&strong&long-range magnetic order for $T < T_{c}$\\
\hline 
Antiferromagnetism&absent&long-range magnetic order for $T < T_{N}$\\
\hline 
Weak ferromagnetism&weak&long-range magnetic order for $T < T_{DM}$\\
\hline
\end{tabular}
\end{center}
\end{table}
The magnetic behavior of materials can be classified into the  six major groups as shown in Table 1.\\
All the materials respond to magnetic fields in essentially different way. 
The simplest magnetic systems to consider are insulators where
electron-electron interactions are weak. If this is the case, the
magnetic response of the solid to an applied field is given by the
sum of the susceptibilities of the individual atoms. The magnetic susceptibility is defined by the the 
2nd derivative of the free energy,
\begin{equation}
\label{2.1}  \chi  =  - \frac{\partial^{2} F}{\partial H^{2} }.
\end{equation}
With the aid of the analysis of the susceptibility one can understand (on the basis of an
understanding of atomic structure) why some systems (e.g. some
elements which are insulators) are paramagnetic ($\chi > 0$) and
some diamagnetic ($\chi < 0$).\\
From the phenomenological point of view magnetic materials are characterized by the intrinsic magnetic
susceptibility
\begin{equation}
\label{2.2}  \chi_{int} = \frac{M}{H_{i}},
\end{equation}
where $M$ is the magnetization of a sample and $H_{i}$ is the internal field. When external magnetic field $H_{ext}$
is applied then the experimental susceptibility $\chi_{exp} \sim M/H_{ext}$ can be written as
\begin{equation}
\label{2.3}  \chi_{exp} \sim \frac{M}{H_{i} + \eta_{d}M} = \frac{\chi_{int}}{1 + \eta_{d}  \chi_{int}},
\end{equation}
where $\eta_{d}$ is the demagnetizing factor. In principle, a ferromagnetic material may have no net magnetic
moment because it consists of magnetic domains.\\
In some materials there is no collective interaction of atomic magnetic moments; they belong to
diamagnetic substances~\cite{dorf61}.
Diamagnetic behavior is characterized by repulsion of a substance out of an applied
magnetic field. This behavior arises from the interaction of the applied magnetic field
with molecular or atomic orbitals containing paired electrons. With the exception of the
hydrogen radical, all atomic or molecular materials exhibit some diamagnetic behavior.
This magnetic behavior is temperature independent, and the strength of the interaction is
roughly proportional to the molecular weight of the material.\\
Diamagnetism is a fundamental property of all matter, although it is usually very weak.
It is due to the non-cooperative behavior of orbiting electrons when exposed to an applied magnetic field. 
Diamagnetic substances are composed of atoms which have no net magnetic moments (ie., all the orbital shells are 
filled and there are no unpaired electrons). However, when exposed to a field, a negative magnetization is produced 
and thus the susceptibility is negative and weak.  It does not depend on the temperature.
Diamagnetic susceptibility $\chi_{d}$ is a part of the total susceptibility
of a material $\chi$. It can be represented in a very approximative form as~\cite{dorf61}
\begin{equation}
\label{2.4} \chi \cong  \chi_{p} + \chi_{d}.
\end{equation}
Here $ \chi_{p}$ is the paramagnetic susceptibility.
Diamagnetism of metallic systems~\cite{vons71,dorf61,cra95,blund01,magmat08,coey09}  is the diamagnetic response of the 
electron gas. The diamagnetic susceptibility is given by~\cite{vons71,dorf61}
\begin{equation}
\label{2.5}  \chi_{d} \sim - \frac{1}{3} \mu_{0} \mu_{B}^{2} D(E_{F})
\end{equation}
and
\begin{equation}
\label{2.6}  \chi_{d} \sim  - \frac{1}{3} \chi_{p}.  
\end{equation}
Here $D(E_{F})$ is the density of states at the Fermi level $E_{F}$; $\mu_{B}$ is the Bohr magneton.
Majority of metallic systems are paramagnetic due to the fact that the (positive) $\chi_{p}$ is three times
larger than the (negative) $\chi_{d}.$
The diamagnetic behavior of various molecules and complex compounds and its competition with paramagnetism is rather diverse.
In some cases this interrelation can be estimated to give~\cite{dorf61}
\begin{equation}
\label{2.7} \overline \chi_{p}   \cong  \frac{2}{3}  \frac{(\Delta \overline \chi_{d})^{2}}{\overline \chi_{d}}.   
\end{equation}
For example, for methane molecule~\cite{dorf61} $\chi_{d} \sim - 13.906 \cdot 10^{-6}$ 
and $\chi_{p} \sim + 0.189 \cdot 10^{-6}$.\\
Paramagnetism~\cite{vons71,dorf61,cra95,blund01,magmat08,coey09} is characterized by the attraction 
of a substance into an applied magnetic
field. This behavior arises as a result of an interaction between the applied magnetic field
and unpaired electrons in atomic or molecular orbitals. Typically, paramagnetic
materials contain one or more unpaired electrons, and the strength of paramagnetic
interactions are temperature dependant. However, some substances exhibit temperature independent paramagnetism 
that arises as a result of a coupling between the
magnetic ground state and non-thermally populated excited states. Temperature independent paramagnetism has been
observed for materials with both paramagnetic and diamagnetic ground states, and it is
usually associated with electrically conducting materials.\\
In paramagnetic materials, some of the atoms or ions  have a net magnetic moment due to unpaired electrons in 
partially filled orbitals. One of the most important atoms with unpaired electrons is iron. However, the 
individual magnetic moments do not interact magnetically, and like diamagnetism, the magnetization is zero 
when the field is removed. In the presence of a field, there is now a partial alignment of the atomic magnetic 
moments in the direction of the field, resulting in a net positive magnetization and positive susceptibility.
Paramagnetism is typically considerably stronger than the  diamagnetism. 
In addition, the efficiency of the field in aligning the moments is opposed by the randomizing effects of temperature. 
This results in a temperature dependent susceptibility, termed by   the 
Curie law~\cite{vons71,dorf61,cra95,bert98,blund01}
\begin{equation}
\label{2.8} \chi \cong  \frac{\mu_{0} N m_{0}^{2}}{3 k_{B} T} = \frac{C}{T}  
\end{equation}
or Curie-Weiss law~\cite{vons71,dorf61,cra95,bert98,blund01,magmat08}
\begin{equation}
\label{2.9}  \frac{1}{ \chi}  \cong  \frac{T - \theta_{p}}{C}, \quad   \theta_{p} \geq T{c}.    
\end{equation}
In summary,   diamagnetism is characterized by negative susceptibility
due to the fact that induced moment opposes applied field. 
Diamagnetic behavior is common for noble gas atoms and alkali halide ions
(e.g., $He, Ne, F^{-}, Cl^{-}, Li^{+}, Na^{+}, \ldots$).
Paramagnetism has positive susceptibility
because induced moments are favored by applied field (but are opposed by thermal disorder).
In paramagnetic substance magnetization is immediately lost upon removal of field.
Paramagnetism is observed for isolated rare earth ions, iron (group $3d$) ions
(e.g., $ Fe^{3+}, Co^{2+}, Ni^{2+}, Sm^{+}, Er^{+}, \ldots$).\\
Ferromagnets and ferrimagnets     are characterized by
strong exchange interaction between localized atomic moments or strong electron correlation among itinerant (narrow band) 
electrons. The interaction arises from the electrostatic electron-electron
interaction, and is called the \emph{exchange interaction}  or exchange force.
Ferromagnetic materials exhibit parallel alignment of moments resulting in large net magnetization even in the 
absence of a magnetic field.
Ferromagnets will tend to stay magnetized to some extent after being subjected to an external magnetic field. 
This tendency to  remember their magnetic history  is called hysteresis~\cite{bert98}. 
The fraction of the saturation magnetization which is retained when the driving field is removed is called the 
remanence of the material, and is an important factor in permanent magnets~\cite{magmat08}.
All ferromagnets have a maximum temperature where the ferromagnetic property disappears as a result of thermal motion. 
This temperature is called the Curie temperature  $T_{c}$.
Ferromagnetic materials are spontaneously magnetized below a temperature $T_{c}$ and all local moments have a 
positive component along the direction of the spontaneous magnetization. In antiferromagnet  individual local moments 
sum to zero total moment (no spontaneous magnetization) whereas in 
ferrimagnet  local moments are not all oriented in the same
direction, but there is a non-zero spontaneous magnetization.\\
In conventional magnetic materials the magnetic ions (magnetic moments)  reside on a regular lattice. The interactions
between moments is generally short-range, determined by overlap of the electron wave functions in conjunction 
with Pauli's exclusion principle. Thus the exchange interaction arises due to the Coulomb electrostatic interaction~\cite{tyab,kuz09}.
The coupling, which is quantum mechanical in nature, is termed as the 
exchange interaction. As a rule only 
nearest neighbor interactions between magnetic moments are essential. 
The exchange interaction between the neighboring magnetic 
ions will force the individual moments into  ferromagnetic parallel 
 or antiferromagnetic  antiparallel  alignment with their neighbouring magnetic moments.  
Thus the important interactions responsible for
ordering and magnetic dynamics in magnetic materials are the strong short-ranged correlations between electrons.
Critical temperatures and magnetic excitation energies are therefore mainly determined by the short range interactions,
and the weak long range dipolar interactions are significant only for long wavelength dynamic behavior and phenomena
related to domain formation. Magnetic dipolar interaction
\begin{equation}\label{2.10}
E_{dd}  \sim      \frac{ \vec{\mu}_{1} \vec{\mu}_{2} - \textrm{3} (\vec{\mu}_{1}\cdot \vec{r}) (\vec{\mu}_{2}\cdot \vec{r})}{r^{3}}              
\end{equation}
is too weak ($E_{dd} \sim 1$ K or $10^{-4} eV$) to account for the ordering of real magnetic materials.\\
There are various types of  the exchange interaction: direct exchange, indirect exchange, superexchange~\cite{vons71,tyab,kuz09,kuz02}.
Direct exchange interaction is effective  for moments, which are close enough to have sufficient overlap of 
their wave functions (atomic-like orbitals). It produces a strong but short range 
coupling which decreases rapidly as the ions are separated. For direct inter-atomic exchange the corresponding integral $J_{\textrm{ex}}$ can be positive 
or negative depending on the balance between the Coulomb interaction and kinetic energy.\\
Indirect exchange interaction~\cite{vons71,tyab,kuz09,kitt68} couples moments over relatively large distances. 
It is the dominant exchange interaction in metals, 
where there is little or no direct overlap between neighboring electrons. It therefore acts through an intermediary, 
which in metals are the conduction electrons (itinerant electrons). This type of exchange is 
termed as the RKKY interaction. This type of interaction is especially relevant for the
rare earth metals with the unfilled $4f$ shell. In these metals the direct exchange between the localized  magnetic
moments on the rare earth ions is negligible, but they are coupled through the medium of the conduction electrons
by the indirect exchange interaction~\cite{vons71,tyab,kuz09,kitt68,elli70,scom00}. This interaction is long range and and oscillatory and, 
together with the strong anisotropy forces, which are a consequence of the anisotropic charge distribution in the $4f$ shell,
lead to the complex magnetic structures.\\
Superexchange describes the interaction between moments on ions too far apart to be connected by direct exchange~\cite{vons71,ander63,marsh66}.
This exchange is relevant, for example, for  ferric-rare earth interaction in   garnets.
In these substances the coupling between the moments on a pair of metal cations 
separated by a diatomic anion is described by the superexchange interaction.
In the P.W. Anderson theory of superexchange~\cite{vons71,ander63,marsh66} a set of magnetic orbitals,
localized at the metal sites, defines the ground and excited state configurations.
The main feature of his approach is that the covalent interaction between the metal
and ligand orbitals is already included from the beginning while the on-site
electron repulsion remains strong enough to keep the electrons mainly localized
on the metals. 
The  ion of  $Fe$  in a garnet has a half filled $3d$ shell and so has a spherically symmetric charge distribution.
The  rare-earth ion is not symmetric and has a strong spin-orbit coupling. As a consequence, its charge distribution is 
coupled to its moment. The ion's moments will be coupled by superexchange due to the ability of the $Fe$ moment alters the 
overlap of the   cation. This will lead to the changing of the magnitude 
of both the Coulomb and exchange interactions between the cations, leading to a coupling, 
which depends on the moment's orientation. The effective perturbation Hamiltonian
reduces in every order of perturbation to the well-known effective spin-Hamiltonian of Heiseberg-Dirac type.\\
Anderson's theory has a few shortcomings. The ground state configurations may not be reasonable starting points 
for perturbation theory, due to the fact that the first-order energy of the singlet state is raised too
much above the triplet energy.  Moreover, it may be inadequate
due to its perturbational description of covalency.\\ 
Transition and rare-earth metals and especially compounds containing transition and
rare-earth elements possess a fairly diverse range of magnetic properties. 
The elements $Fe$, $Ni$, and $Co$ and many of their alloys are typical ferromagnetic materials.
The construction of a consistent microscopic theory explaining the magnetic properties of these substances encounters
serious difficulties when trying to describe the collectivization-localization duality in the behavior
of magneto-active electrons~\cite{kuz09,kuz81,akuz02,kuz02}. This problem appears to be extremely important, since its solution
gives us a key to understanding magnetic, electronic, and other properties of this diverse group of
substances. Quantum theory of magnetism deals with variety of the schematic models of magnetic
behavior of real magnetic materials. In papers~\cite{kuz09,akuz02} we presented a comparative analysis of these
models; in particular, we compared their applicability for description of complex magnetic
materials.\\  
Magnetic properties and the dynamic response of magnetic systems are strongly dependent on 
dimensionality~\cite{kuz10} and on size~\cite{cheon06,may09}. Saturation moments $M_{S}$ of magnetic materials depart from their bulk values near
a surface because of reduced symmetry and altered charge distribution, which is typical variation over a few Angstroms
in metals. Another reason is the surface stress and/or surface segregation (about  several tens of Angstroms).
Surface magnetic effects are evident in studies of thin magnetic films 
and multilayers~\cite{hei93,sco03,scom06,mil06,zab10,cheon06,getz10}. The intrinsic magnetic 
properties - magnetization, critical temperature, anisotropy, magnetostriction - may differ substantially in thin
films and bulk material. It is worth noting that sometimes the substrate influences the electronic structure and
magnetic moment of the first atomic layers at the interface. Magnetic films with thickness ranging from a single
monolayer to a few monolayers and bigger (up to $\sim 100$nm) may be grown on crystalline or amorphous substrates.
In the transition (3$d$)  metal  films surface atoms have less of their neighbors. Thus the exchange interactions
may be more weak. In terms of the itinerant picture the corresponding bands may be more narrow at the surface. As a result 
the local density of states and the  local magnetic moments may be enhanced. These effects as a rule are limited
to the first one or two monolayers. \\
Magnetic multilayers~\cite{bruno1,bruno2,bruno3} typically consist of alternate stacks of ferromagnetic and nonferromagnetic
spacer layers. The typical thickness of an individual layer ranges between
a few atomic layers   to a few tens of atomic layers. The magnetic layers usually
consist of elemental metallic ferromagnets $(Fe, Co, Ni)$ or alloys thereof (e. g. permalloy).
The spacer layers can consist of any transition or noble metal; they are either
paramagnetic $(Cu, Ag, Au, Ru, Pd, V, etc.)$ or antiferromagnetic $(Cr, Mn)$.
Because of the spacer layers, the magnetic layers are, to first approximation, magnetically
decoupled from each other, i. e. their basic magnetic properties such as
magnetization, Curie temperature, magnetocrystalline anisotropy, magneto-optical
response, \emph{etc.}, are essentially those of an individual layer. This approximation, however,
is not sufficient for accurate description of the magnetism of multilayers, and
one must consider the magnetic interactions which couple successive magnetic layers
through spacer layers.\\
The  interactions  which give rise to an interlayer magnetic interaction are essentially
the dipolar interaction and   the indirect exchange interaction of the Ruderman-
Kittel-Kasuya-Yosida (RKKY) type~\cite{vons71,tyab,kuz09,kitt68,elli70,scom00}.
For a homogeneously magnetized layer consisting of a continuous medium, there
is no dipolar stray field, so that dipolar interlayer coupling can arise only as a result of
departures from this idealized situation. This is the case when one considers the real
crystalline structure of the layer. However, the dipolar stray
field decays exponentially as a function of the distance from the magnetic layer, with
a decay length of the order of the lattice parameter, so that this effect is completely
negligible compared with the interaction as a result of exchange~\cite{bruno1,bruno2,bruno3}.\\
The indirect exchange interaction has a completely different physical origin. It
is mediated by conduction electrons which are scattered successively by the magnetic layers.
In metallic systems, exchange interactions are mediated by itinerant electrons
and thus can be transmitted over relatively long distances. It follows that
exchange interactions can couple magnetic layers through non-magnetic metallic
layers.  The interest of the
exchange coupled multilayers has also been enhanced by the discovery of the
giant magnetoresistance~\cite{bruno1,bruno2,bruno3}. \\
In order to understand the intrinsic properties of nanomagnets, much effort has been devoted to $3d$ ferromagnetic transition-metal
clusters~\cite{rod06,kubl03,emil06,erodun06} such as $Fe$, $Ni$, $Co$ and $Pd$, $Pt$  and rare earth $4f$ aggregates~\cite{hirt99} 
such as $Gd$ and $Tb$. These studies provide insight into the electronic structure of the cluster and is fundamental for an understanding
of how magnetism develops in small cluster. \\
Intensive research on fullerenes, nanoparticles, and quantum dots  led to interest in clusters, fullerenes, nanotubes 
and nanowires in last decades~\cite{handnano10,hnano10,hnano10a}. The studies of nanophysics of materials, in particular clusters, fullerenes, nanotubes and nanowires  posed
many important problems of condensed matter physics in these  nanoscale materials and structures.\\
It should be stressed that
nanophysics~\cite{rod06,emil06,fuller93,fuller11,carb3,handnano10,hnano10} brings together multiple disciplines to determine the structural, electronic, optical, 
and thermal behavior of nanomaterials. It includes also the electrical and thermal conductivity, the forces between 
nanoscale objects  and the transition between classical and quantum behavior. These features are also the key aspects 
of carbon nanotubes, including quantum and electron transport, isotope engineering, and fluid flow, which are relevant
also for inorganic nanotubes, such as spinel oxide nanotubes, magnetic nanotubes, and self-assembled peptide nanostructures. 
%
%
%
%
%
\section{Microscopic  Models of Magnetic Substances}
%
It is instructive to make a quick overview of the microscopic basis of the
quantum physics of magnetism~\cite{tyab,kuz09}.
It is well known that the quantum mechanics is the key
to understanding magnetism. One of the first
steps in this direction was the formulation of  Hund's
rules  in atomic physics. 
In  rare-earth and transition metals elements  the atomic shells are partially filled, and
the ground state is determined by minimizing the atomic energy together with the intra-atomic Coulomb interaction needed
to remove certain degeneracies~\cite{tyab,kuz09,kuz10}.
A physical analysis of the first Hund's
rule leads us to the conclusion, that it is based on the fact,
that the elements of the diagonal matrix of the electron-electron's Coulomb interaction contain the exchange's
interaction terms, which are entirely negative. This is the
case only for electrons with parallel spins. Therefore, the
more electrons with parallel spins involved, the greater
the negative contribution of the exchange to the diagonal
elements of the energy matrix. Thus, the first Hund's rule
implies that electrons with parallel spins tend to avoid each other spatially. 
Here, we have a direct connection
between Hund's rules and the Pauli exclusion principle, which states that 
two electrons in the same orbit must move in opposite directions.\\
Thus it should be stressed once again that the origin of strong ferromagnetism is
the electron-electron's Coulomb interaction $V (r_{i} - r_{j}) \sim e^{2}/(r_{i} - r_{j})$. 
The corresponding energy scales are presented in Table 2.  
%
%
%
\begin{table}
\begin{center}
Table 2. Energy scales for magnetic ordering
\end{center}
\begin{center}
\begin{tabular}{|l|l|} \hline
Energy scale&Order of magnitude \\
 \hline \hline
Coulomb interaction & $\simeq$ eV\\
 \hline
Magnetic ordering temperature& $T_{c} \sim$  0.1 meV - 100 meV \\
 \hline
RKKY interaction& $\sim$ 0.1 meV  - 1 meV\\ 
 \hline
dipole-dipole interaction & $\sim$  0.1 meV\\
 \hline
\hline
\end{tabular}
\end{center}
\end{table}
%
%
%
%
A schematic realization of of the above principles gives
the method of model Hamiltonians  which has proved to be
very efficient in the theory of magnetism~\cite{tyab,kuz09}. Without any
exaggeration one can say, that the tremendous successes
in the physics of magnetic phenomena were
achieved, largely, as a result of exploiting a few simple
and schematic model concepts for "the theoretical
interpretation of ferromagnetism"~\cite{kuz09,kuz02,rscom08}.\\ 
One can regard (with some reservation)
the Ising model~\cite{vons71,tyab,kuz09} as the first model of the quantum
theory of magnetism. In this model  it was assumed that the spins are
arranged at the sites of a regular one-dimensional lattice.
Each spin $S^{z}$  can obtain the values $\pm \hbar/2$:
\begin{equation}\label{3.1}
\mathcal{H}  \sim  - \sum_{<ij>}  I_{ij} S^{z}_{i}S^{z}_{j}.
\end{equation}
Here $I_{ij}$ is the  parameter of the spin-spin interaction.
This was one of the first attempts to describe the
magnetism as a cooperative effect. \\ 
However, the Ising model oversimplifies the situation in real
crystals. Works of W. Heisenberg, P. Dirac  and van Vleck have lead to devising of a more
general theory which attributed  the ferromagnetic state to an alignment of electron spins
in atoms due to exchange forces.
The Heisenberg model 
describes  schematically the interaction between spins at different sites
of a lattice by the following isotropic scalar function
\begin{equation}\label{3.2}
\mathcal{H}  = -  \sum_{ij}  J(i-j)   \textbf{S}_{i}  \textbf{S}_{j} 
-g\mu_{B}H_{ext}  \sum_{i}S_{i}^{z}. 
\end{equation}
Here $\textbf{S}_{i}$ is the spin angular momentum operator of the atom at site $i$;
$\mu_{B}$ is the Bohr magneton and $g$ is the gyromagnetic factor 
(the magnetic moment $\mu_{0}$ is defined as  $\mu_{0} = 1/2  g   \mu_{B}$); $H_{ext}$ is the intensity
of a static magnetic field directed along the $z$ axis; for the case $H_{ext} > 0$ the magnetic moments
line up along the positive $z$ axis when the system is in the ground state.
The quantity $J(i-j)$  (the "exchange integral") is the strength of
the exchange interaction between the spins~\cite{tyab,rscom08,vv53,herrin66} located at the
lattice sites $i$ and $j$. The exchange force is a quantum mechanical phenomenon due to the 
relative orientation of the spins of two electron. 
Exchange force depends on relative orientation of spins of two electrons due to Pauli's exclusion principle.
It is usually assumed that $ J(i-j) = J(j-i)$  and $J(i - j = 0) = 0,$  which means that only the
inter-site interaction is present (there is no self-interaction).\\ 
The Heisenberg Hamiltonian (\ref{3.2}) can be rewritten
in the following form:
\begin{equation}
\label{3.3} \mathcal{H} = -  \sum_{ij} J(i-j) ( S^z_{i}S^z_{j} +
 S^+_{i}S^-_{j}).
\end{equation}
Here, $S^{\pm} = S^x \pm iS^y$ are the spin raising and lowering
operators. 
Note that in the isotropic Heisenberg model the $z$-component
of the total spin $S^z_{tot} = \sum_{i}S^z_{i}$ is a constant of
motion, that is  $ [H,S^z_{tot}] = 0$.\\
Exchange forces are very large, equivalent to a field on the order of $ \sim 10^{3}$ Tesla.
In real substances direct exchange is driven by minimizing potential
energy, by reducing wave functions overlap. To clarify this let us consider the Coulomb interaction $V$.
For two-electron antisymmetric wave function  
\begin{equation}
\label{3.4} \psi (r_{i},\sigma; r_{j} ,\sigma)  =  
\frac{1}{\sqrt{2}} \Bigl ( \varphi_{\alpha}(r_{i})\varphi_{\beta}(r_{j}) - \varphi_{\beta}(r_{i})\varphi_{\alpha}(r_{j}) \Bigr )
\chi_{\sigma} \chi_{\sigma},
\end{equation}
matrix element $\langle \psi |V|  \psi \rangle$  will take the form
\begin{eqnarray}
\label{3.5} 
E_{C} \sim  \langle \psi(i,j) |V|  \psi(i,j) \rangle = \\ \nonumber
\int d^{3}r_{i} d^{3}r_{j} V(r_{i} - r_{j}) |\varphi_{\alpha}(r_{i})|^{2} |\varphi_{\beta}(r_{j})|^{2} - \\ \nonumber
\int d^{3}r_{i} d^{3}r_{j} V(r_{i} - r_{j})  \varphi_{\alpha}^{*}(r_{i})   \varphi_{\beta}^{*}(r_{j}) \varphi_{\alpha}(r_{j})\varphi_{\beta}(r_{i}) \\ \nonumber  = 
U_{\alpha \beta} - J_{\alpha \beta}. 
\end{eqnarray}
Here $U$ is the Coulomb interaction energy and $J$ is the effective exchange integral.
In real material  the calculation of   the effective exchange integral 
$J_{\alpha \beta}$ is very difficult task~\cite{kuz09,her66,herrin66,ander63}; it
depends substantially on the proper choice of
the many-electron wave functions for the system. In the work by Mattheiss~\cite{mat61} the magnetic properties of 
a linear chain of monovalent atoms were investigated from the point of view  of perturbation theory. 
The many-electron wave functions for the system were expanded as linear combinations of determinantal functions which were
eigenfunctions of  $S^{2}$  and  $S^z_{tot}.$ These determinantal functions were constructed  from orthonormal 
one-electron orbitals of the Wannier type so that the nearest neighbor exchange integral is positive definite and approaches zero at large 
lattice spacing. The analytic expression was found for an effective nearest neighbor exchange integral $J$.
The main conclusion which follows from this consideration is that the Heisenberg type model is not a reasonable workable model
at all internuclear separations. Rather, it implies that there is a gradual transition from the energy band to the
orthonalized atomic orbitals approximations which are valid at small and large lattice spacing respectively. The magnetic
interaction is not \textbf{exactly} described by the Heisenberg exchange operator.\\
Thus, in the framework of the Heisenberg-Dirac-van Vleck model~\cite{vons71,tyab,kuz09}, describing the interaction
of localized spins, the necessary conditions for the
existence of ferromagnetism involve the following two
factors. Atoms of a "ferromagnet to be" must have a
magnetic moment, arising due to unfilled electron $d$- or
$f$-shells. This is related simply to the fact that both the $3d$ and $4f$
wave functions are strongly localized and have no nodes in radial wave functions.
Contrary to this the $4p$ and $4s$ wave functions are  delocalized and have  nodes in radial wave functions.
The exchange integral $ J_{ij}$ related to the electron
exchange between neighboring atoms must be positive.
Upon fulfillment of these conditions the most energetically
favorable configurations in the absence of an
external magnetic field correspond to parallel alignment
of magnetic moments of atoms in small areas of
the sample (domains). \\ Of course, this simplified
picture is  a scheme   only. A detail derivation of the
Heisenberg-Dirac-van Vleck model describing the
interaction of localized spins is quite complicated~\cite{kuz09,her66,herrin66,ander63}.
An important point to keep in mind here is that magnetic
properties of substances are born by quantum effects,
the forces of exchange interaction.\\
As was already mentioned above, the states with
antiparallel alignment of neighboring atomic magnetic
moments are realized in a fairly wide class of substances.
As a rule, these are various compounds of transition
and rare-earth elements, where the exchange
integral $ J_{ij}$ for neighboring atoms is negative. Such a
magnetically ordered state is called 
\textbf{antiferromagnetism}~\cite{vons71,tyab,kuz09} which was explained by L. Neel.
For example, the transition-metal compound $MnF_{2}$ (manganese fluoride)
have the antiferromagnetic behavior at low temperatures ($T_{N} \sim 66 K$).
In this compound $Mn$ ion becomes  $Mn^{2+}$ ($3d^{5}$ configuration) and fluorine ion - $F^{-}$.\\ 
In 1948, L. Neel introduced also the notion
of \textbf{ferrimagnetism}~\cite{vons71,tyab,kuz09} to describe the properties
of substances in which spontaneous magnetization
appears below a certain critical temperature due to nonparallel
alignment of the atomic magnetic moments.
These substances differ from antiferromagnets
where sublattice magnetizations $m_{A}$ and $m_{B}$
usually have identical absolute values, but opposite orientations.
Therefore, the sublattice magnetizations
compensate for each other and do not result in a macroscopically
observable value for magnetization. In ferrimagnetics
the magnetic atoms occupying the sites in
sublattices $A$ and $B$  differ both in the type and in the
number. Therefore, although the magnetizations in the
sublattices $A$ and $B$ are antiparallel to each other, there
exists a macroscopic overall spontaneous magnetization.
The antiferromagnetics and ferrimagnetics constitute a very wide group
of various substances. \\
Later, substances possessing weak ferromagnetism
were investigated~\cite{vons71,tyab,kuz09}. It is interesting that originally
Neel used the term \emph{parasitic ferromagnetism} 
when referring to a small ferromagnetic moment,
which was superimposed on a typical antiferromagnetic
state of the $\alpha-$iron oxide $Fe_{2}O_{3}$ (hematite).
Later, this phenomenon was called canted antiferromagnetism,
or weak ferromagnetism. The
weak ferromagnetism appears due to antisymmetric
interaction between the spins $\vec{S}_{1}$ and $\vec{S}_{2}$ and which is proportional
to the vector product $\vec{S}_{1} \times \vec{S}_{2}.$ This interaction
is written in the following form
\begin{equation}\label{3.6}
\mathcal{H}_{DM}  \sim  \vec{D}   \vec{S}_{1} \times \vec{S}_{2}.
\end{equation}
The interaction (\ref{3.6}) is called the Dzyaloshinsky-Moriya interaction~\cite{vons71,tyab,kuz09,chu12}. \\
Thus, there exist a large number of substances and
materials that possess different types of magnetic
behavior: diamagnetism, paramagnetism, ferromagnetism,
antiferromagnetism, ferrimagnetism, and weak
ferromagnetism. We would like to note that the variety
of magnetism is not exhausted by the above types of
magnetic behavior; the complete list of magnetism
types is substantially longer~\cite{hurd}. As was already
stressed, many aspects of this behavior can be reasonably
well described in the framework of a very crude
Heisenberg-Dirac-van Vleck model of localized spins.
This model, however, admits various modifications. Therefore, various
nontrivial generalizations of the localized spin models
were studied~\cite{vons71,tyab,kuz09}.\\ 
The Heisenberg model describing localized spins is
mostly applicable to substances where the ground
state's energy is separated from the energies of excited
current-type states by a gap of a finite width. That is, the
model is mostly applicable to semiconductors and
dielectrics. However, the main strongly magnetic
substances, nickel, iron, and cobalt, are metals,
belonging to the transition group~\cite{vons71,tyab,kuz09}.  In many cases inter-electron interaction
is very strong and the description in terms of the
conventional band theory is no longer applicable. Special
properties of transition  metals and of their
alloys and compounds are largely determined by the
dominant role of $d$-electrons~\cite{kuz09,her66,kubl,mizia}. In contrast to simple metals,
where one can apply the approximation of quasi-free
electrons, the wave functions of $d$-electrons are
much more localized, and, as a rule, have to be
described by the tight-binding approximation. The main aim of the band theory of magnetism
is to describe in the framework of a unified approach both
the phenomena revealing the localized character of
magnetically active electrons, and the phenomena
where electrons behave as collectivized band entities~\cite{kuz09}.\\
The quantum statistical
theory of systems with strong inter-electron correlations
began to develop intensively when the main features
of early semi-phenomenological theories were formulated
in the language of simple model Hamiltonians~\cite{kuz09,her66,kubl,mizia,rscom08}. 
The most known models are the Anderson model~\cite{kuz96} and Hubbard model~\cite{akuz02,kuz02}.
Both the Anderson model~\cite{kuz09},  
and the Hubbard model~\cite{kuz09,kuz02}  equally stress the role
of inter-electron correlations. The Hubbard Hamiltonian
and the Anderson Hamiltonian (which can be considered
as the local version of the Hubbard Hamiltonian)
play an important role in the electron's solid-state theory~\cite{kuz09,akuz02,kuz02}.\\
The Hamiltonian of the Hubbard model is
given by:
\begin{equation}\label{3.7}
\mathcal{H} = \sum_{ij\sigma}t_{ij}a^{\dagger
}_{i\sigma}a_{j\sigma} + U/2\sum_{i\sigma}n_{i\sigma}n_{i-\sigma}.
\end{equation}
The above Hamiltonian includes  the
one-site intra-atomic Coulomb repulsion $U$, and $t_{ij}$, the one-electron
hopping energy describing jumps from a $j$ site
to an $i$ site. As a consequence of correlations electrons
tend to "avoid one another". Their states are best modeled~\cite{kuz09,akuz02,kuz02}
by orthonormal atomic-like Wannier wave functions $[\phi({\vec r} -{\vec R_{j}})]$.
The Hubbard model's Hamiltonian can be characterized
by two main parameters: $U$, and the effective band
width of tightly bound electrons
$$\Delta = (N^{-1}\sum_{ij}\vert t_{ij}\vert^{2})^{1/2}.$$
The band energy of Bloch electrons $\epsilon(\vec k)$ is given by
$$\epsilon(\vec k)     = N^{-1}\sum_{\vec k}t_{ij} \exp[- i{\vec
k}({\vec R_{i}} -{\vec R_{j}}],$$
where $N$ is the total number of lattice sites. Variations
of the parameter $\gamma = \Delta/U$ allow one to study two interesting
limiting cases, the band regime ($\gamma \gg 1$) and the
atomic regime ($\gamma \rightarrow 0$).
Note that the single-band Hubbard model (\ref{3.7}) is a
particular case of a more general model, which takes
into account the degeneracy of $d$-electrons. 
It is necessary to stress that the Hubbard model is
most closely connected with the \textbf{Pauli exclusion principle},
which prohibits the double occupancy  at the same site.
In this case it can be written as $n^{2}_{i\sigma} = n_{i\sigma}$. \\
A generalized spin-fermion model, which is also
called the Zener model, or the $s$-$d$- ($d$-$f$)-model is of
importance in the solid-state theory. The Hamiltonian
of the $s$-$d$ exchange model~\cite{vons71,tyab,kuz99,kuz04,kuz05} is given by:
\begin{equation}\label{3.8}
\mathcal{H} = H_{s} + H_{s-d},
\end{equation}
\begin{equation}\label{3.9}
 H_{s} = \sum_{k\sigma}\epsilon_{k}c^{\dag}_{k\sigma}c_{k\sigma},
\end{equation}
\begin{equation} \label{3.10}
H_{s-d} = J {\vec \sigma_{i}}{\vec S_{i}} = - J N^{-1/2}\sum_{kk'} 
\left (c^{\dag}_{k' \uparrow}c_{k \downarrow}S^{-}  + c^{\dag}_{k' \downarrow}c_{k \uparrow}S^{+} +
(c^{\dag}_{k' \uparrow}c_{k \uparrow} - c^{\dag}_{k' \downarrow}c_{k \downarrow})S^{z}  \right ).
\end{equation}
Here, $c^{\dag}_{k\sigma}$ and $c_{k\sigma}$ are the second-quantized operators
creating and annihilating conduction electrons. The
Hamiltonian (\ref{3.10}) describes the interaction of the localized
spin of an impurity atom with a subsystem of
the host-metal conduction's electrons. This model is
used for description of the Kondo effect and other problems.
Note that the Hamiltonian of the $s$-$d$ model is a low-energy realization of the Anderson model.\\
Thus, the Anderson,  Hubbard  and $s$-$d$-   models take into account
both the collectivized (band) and the localized behavior
of electrons~\cite{kuz09,akuz02,kuz02,kuz99,kuz04,kuz05} whereas Heisenberg model emphasizes the localized character
of magneto-active electrons (see Table 3).  
%
%
%
%
%
\begin{table}
\label{tab3}
\begin{center}
Table 3. Magnetic interactions: localized and itinerant models
\end{center}
\begin{center}
\begin{tabular}{|l|l|l|} \hline
Interaction&Hamiltonian&Curie Temperature\\
 \hline
1: Dipolar  interaction&$E_{dd} \sim \mu^{2}/4 \pi r^{3}$&$E \sim 1 K$\\ 
 \hline
2: Exchange direct&$H \sim \sum J_{ij}S_{i}S_{j}$&$T_{c} \sim zJ \sim 10 - 10^{3} K$\\
 \hline
3: Exchange indirect&$H \sim  I(r) S_{i} S_{j}$&$T_{c} \sim 1  - 10^{2} K$\\
 \hline
4: RKKY&$I(r) \sim \cos (2k_{F}r)/r^{3}$&$T_{c} \sim 1  - 10^{2} K$\\
 \hline
5: Itinerant electron magnet&$H \sim  U \sum n_{i\uparrow}n_{i\downarrow}$&$T_{c}\sim 10 - 10^{3} K$\\
 \hline
\end{tabular}
\end{center}
\end{table}
%
%
%
\\
To summarize the studies of the previous sections, the term magnetism refers to substances that at the atomic level exhibit
temperature dependent paramagnetic behavior as the most characteristic feature~\cite{vons71,hurd}. The
non-zero spin angular moment associated with an unpaired electron gives rises to a
magnetic moment.  As a rule, in condensed matter, bulk magnetic properties
arise as a result of long-distance interactions between atomic-like electrons.\\
The control of bulk magnetic properties has proven to be very difficult~\cite{magmat08,stef08,stef12,spald11,yzhu}. 
This is also due to the many different types of magnetic behaviors that should 
be  characterized and identified~\cite{hurd,yzhu}. These
different magnetic behaviors arise from the various types of electron and spin interactions observed in
these materials leading to variety of different models for their interpretation~\cite{kuz09,rscom08}. 
The temperature of magnetic ordering can be varied in a very broad interval
of temperatures. Table 4 provides examples of
different magnetic materials with big variety of 
ordering temperature~\cite{magmat08,stef08,coey09,spald11,stef12,hand05,crc09}.
The complexity associated 
with controlling magnetic properties has
arisen from difficulties in controlling the spatial arrangement of spin containing units~\cite{magmat08,stef08,stef12,spald11,yzhu}.
There are  several strategies
for controlling spin-spin organization; the most efficient one of these strategies is the neutron
scattering technique~\cite{kuz81,yzhu,ml71,tapa06}.  \\
Traditional magnetic materials are two- and three-dimensional arrays of inorganic atoms,
composed of transition metal or lanthanide metal containing spin units. These materials
are typically produced at very high temperatures using the best achievements of metallurgical sciences. In
contrast to traditional magnetic materials, molecular and carbon-based materials are organic or
hybrid (inorganic/organic)  materials. Certain fraction of  these materials may manifest some magnetic features. 
They comprised of either metal containing spin units or
organic radical containing spin units. It has been conjectured that these materials will
allow for the low temperature synthesis of magnetic materials, materials with better
optical properties, the combination of magnetic properties with mechanical, electrical,
and/or optical properties. In addition, they may provide a better control over a material's magnetic 
characteristics~\cite{magmat08,stef08,stef12,spald11,yzhu,handnano10,hnano10}.
However, in order to design materials with interesting bulk magnetic properties it is necessary to
understand how bulk magnetism arises in samples. A complete resolution  
of this task has not yet been achieved in the full measure~\cite{lou06}, however many of the general principles of
magnetic behavior are well established~\cite{rod06,sieg06,magmat08,stef08,coey09,spald11,stef12,kuz09}. Some of those principles may be used in   discussion of the
complicated problem of the possible magnetic properties of carbon-based structures~\cite{kuz11,kuz12,kuz12a}.
%
%
%
\begin{table}
\label{tab4}
\begin{center}
Table 4. Magnetic properties of various compounds~\cite{magmat08,stef08,coey09,spald11,stef12,hand05,crc09}
\end{center}
\begin{center}
\begin{tabular}{|l|l|l|} \hline
Compound&Type of ordering&Critical Temperature\\
 \hline
$Mn_{3}N_{2}$&AFM&$  T_{N} \sim 925 K$\\
\hline
$Ni_{3}Fe$&FM&$  T_{c} \sim 620 K$\\
 \hline
$FeS$&AFM&$   T_{N} \sim 305 K$\\  
 \hline
$CrN$&AFM&$  T_{N} \sim 273 K$\\
 \hline
$MnSi$&AFM&$  T_{N} \sim 30 K$\\
 \hline
$ZrZ_{2}$&weak FM&$T_{c} \sim 28 K$\\
 \hline
$ZnCu_{3}(OH)_{6}Cl_{2}$&spin-1/2 kagome-lattice AFM&$  T_{N} \sim 0.05 K$\\
 \hline
\end{tabular}
\end{center}
\end{table}
%
%
%
%
%
%
\section{Carbon and its Allotropes}
%
%
Carbon materials are unique in many ways~\cite{chem95,carb1,fuller93,fuller11,carb2,carb3}. They are characterized by the 
various allotropic forms that carbon materials can assume~\cite{cage88,praw97,hirsh10}. 
Chemical properties of carbon are remarkably versatile. Carbon electronic structure is $1s^{2}2s^{2}2p^{2}$.
Moreover, the outer shell $2s$ and $2p$ electrons can hybridize in triple ways, forming $sp^{1}$, $sp^{2}$
and $sp^{3}$ orbital wave functions. Each type of hybridization is realized in a material with substantially
different properties. The $sp^{1}$ hybridization provides the formation of $2\sigma$  and $2\pi$ orbitals. 
Thus this hybridization  favours linear structures such as those observed in polymers.
The $sp^{2}$ hybridization results in three strong $\sigma$ bonds, with an unhybridized $p$ electron forming a $\pi$ bond.
This case is appropriate for description of graphite and and graphite-like materials with planar two-dimensional structures.
The $sp^{3}$ hybridization leads to four identical $\sigma$ bonds arranged tetrahedrally in three dimensions.
This case is suitable for the strong bonding in crystalline diamond. Diamond is a nonconductor.
Thus different bonding results in radically
different physical properties. For systems consisting of a mixture of bonding types (e.g. evaporated carbon, tetrahedral
amorphous carbon, glassy carbon, etc.)  a dominant factor in determining their properties is the proportion 
of the carbon atoms which are four-fold coordinated, i.e. the so called $sp^{3}$ content.\\
The materials science of carbon has  been a rich
area of discovery and development in the past decades. Until
relatively recently, the only known polymorphs of carbon
were graphite, in which $sp^{2}$ hybridized carbon atoms form
planar sheets in a two-layer hexagonal stacking, and diamond,
in which $sp^{3}$ carbons form a three-dimensional framework
of cubic symmetry. 
In graphite the three $sp^{2}$ hybrid orbitals of carbon
form $\sigma-$bonds with its neighbors and establish the
hexagonal lattice (two dimensional honeycomb plane). These
$\sigma-$bonds are  strong  enough. The additional $2p$ orbital of each carbon atom forms $\pi$-bond
with its neighbors. The sideway overlapping of $2p$ orbitals between neighboring
carbon atoms form a diffusive distribution of electrons.
This results in delocalization of electrons within the
honeycomb plane and graphite is thus conducting. Its electrical conductivity is highly anisotropic.
Graphite is metallic when current is flowing within the honeycomb plane and is
semiconducting when current is flowing perpendicular to the
honeycomb plane. From the point of view of his conduction properties graphite should be classified as semi-metal.\\
In graphene sheet the conduction and valence bands
consisting of $\pi$ orbitals cross at $K$ and $K'$ points of the
Brillouin zone, where the Fermi level is located. Graphene is a reasonable good conductor~\cite{oleg10}.\\ 
The  properties  of graphene  can be  modified significantly
by introducing defects and by saturating with hydrogen. 
Graphane is a two-dimensional polymer of carbon and hydrogen with the formula unit $(CH)_{n}$ where $n$ is large.
It can be considered as   a two dimensional analog of cubic diamond.
The carbon bonds of graphane are in $sp^{3}$ configuration, as 
opposed to graphene's $sp^{2}$ bond configuration. \\ 
So far, the only two-dimensional carbon allotrope that can be produced
or synthesized is graphene. However, as it was noted in Ref.~\cite{malko12}  in principle,
infinitely many other two-dimensional periodic carbon allotropes, e.g.,
graphynes or graphdiynes, can be envisioned.
Unlike graphene, which has single or double bonds,
graphynes and graphdiynes may be built from triple- and
double-bonded units of two carbon atoms and it is not restricted to just a hexagonal pattern. 
It was conjectured that the number of patterns that it can exist  may be very big.\\
Thus, as it was long known, the different allotropic forms of carbon have essentially
different structures. Due to the different bonding characteristics they have different chemical and physical
properties  (see Table 5). 
%
%
%
%
%
\begin{table}
\label{tab5}
\begin{center}
Table 5. Carbon-based materials 
\end{center}
\begin{center}
\begin{tabular}{|l|l|l|} \hline
Material&Structure&Bonding\\
 \hline  \hline
Diamond&3D crystal&involving $sp^3$ hybridization\\
\hline
Graphite&3D crystal&involving $sp^2$ hybridization\\
 \hline
Graphene&2D single graphite layer&involving $sp^2$ hybridization\\
 \hline
Graphane&2D polymer&involving $sp^3$ hybridization\\
 \hline  
Carbolite&chain-like crystal&involving $sp^1$ hybridization\\  
 \hline
$C_{60}$& molecule&involving $sp^2$ hybridization\\
 \hline
Carbon nanotube&-& -\\
 \hline
Activated carbon&-& -\\
 \hline
\end{tabular}
\end{center}
\end{table}
%
%
%
%
\\
During the last twenty years, the multiplicity of potential carbon structures has consistently posed a big 
challenge to theoretical and computational physicists. Several different methods are currently being used to study 
the structure and the properties of such systems~\cite{rod06,emil06,novcarb10,sheka12}. These methods include simulations based on empirical 
potentials, tight-binding calculations and density functional theory. A combination of these methods is 
needed to make significant progress in the field of carbon-based structures, forming regular solids and clusters.
Cluster-based solids~\cite{handnano10} illustrate the importance of local order
in determining global properties in solids. These solids
add a new dimension to material science. The stronger intra-cluster
bonding in these materials allows them to keep their
individual identity while forming part of the bulk material.\\
The most striking example of this is the $C_{60}$ molecule~\cite{fuller93}. 
Bulk carbon crystals made of $C_{60}$ are highly stable and form
a metastable phase of carbon in addition to the energetically
most stable bulk graphite. 
Fullerene $C_{60}$ is a molecular crystal formed by all-carbon molecules with a closed-cage structure and a
nearly spherical shape.
$C_{60}$ molecules form a face-centered cubic   crystal lattice at room temperature with
weak van der Waals type bonding between the molecules. The solid forms of other fullerenes can also be expected
to be stable when synthesized under optimal conditions. Indeed, successful synthesis of a solid form composed
of $C_{36}$ fullerene molecules has been reported~\cite{carb36}.\\
The study of transformations of $C_{60}$  fullerene at high pressures and temperatures has
shown that the identification of the different carbon states formed in the system presents
certain difficulties.
V. A. Davydov and his group~\cite{dav96,dav97,dav97a,dav98,dav00,dav04} have achieved big progress 
in the studies of transformations of $C_{60}$ fullerene at high pressures and temperatures.
In Ref.~\cite{dav96} they proposed a scheme for classifying the carbon states
that form under nonhydrostatic compression of $C_{60}$ fullerite at pressures up to $10 \,  GPa$ and
temperatures up to $1900 \, K$. Using the character of the structure-forming element (atom,
molecule, polymolecular cluster) as a criterion, different types of carbon states were
distinguished in the system: molecular, polymolecular (polymerized and polycondensed),
and atomic. They performed an  $x$-ray phase analysis of the polymerized states  on the
basis of the phase identification made in Ref.~\cite{nunez95} 
It was established also by $x$-ray diffraction and Raman scattering that the
polymerization of $C_{60}$ fullerene at $1.5 \, GPa$ and $723 \, K$ leads to the
formation of an orthorhombic phase that is different from the previously
identified high-pressure orthorhombic phase.  The mechanisms leading to the formation
of the polymerized phases was discussed on the basis of the results obtained. 
Thus in the works by V. A. Davydov and his group~\cite{dav96,dav97,dav97a,dav98,dav99,dav00,dav01,dav04} the experimental
and model computational results  of big importance were obtained that made it possible to refine the previously
obtained data on the identification of the polymerized states.\\
As it was shown in Ref.~\cite{nunez95} the heating under high pressure drives $C_{60}$ to new distorted crystalline phases 
that are metastable at room temperature and pressure.
Additional information was obtained in Ref.~\cite{lyap03} where it was shown
that application of nonhydrostatic pressure to cluster-based molecular material, like fullerite $C_{60}$, provides an 
opportunity to create elastically and structurally anisotropic carbon materials, including two-dimensional polymerized 
rhombohedral $C_{60}$ and superhard 
graphite-type $(sp^2)$ disordered atomic-based phases. There is direct correlation between textured polymerized 
and/or textured covalent structure and anisotropic elasticity. Whereas this anisotropy was induced by the uniaxial 
pressure component, in the case of disordered atomic-based phases, it may be governed by the uniform pressure magnitude.\\
In recent paper~\cite{mel12} by K. P. Meletov   and G. A. Kourouklis, 
the great advantages of the $C_{60}$ molecule and its potential for polymerization  
due to which the molecule can be the building block of new all carbon materials were reviewed. 
This substance contains, both $(sp^2)$ and $(sp^3)$ hybridized carbon atoms, which allows synthesizing new carbon 
materials with desired physicochemical properties using both types of carbon bonding. The one- and two-dimensional 
polymeric phases of $C_{60}$ are prototype materials of this sort. Their properties, especially polymerization under 
pressure and room temperature via covalent bonding between molecules belonging to adjacent polymeric chains or polymeric 
layers, can be used for further development of new materials. The  review~\cite{mel12} was focused on the study of the pressure-induced
polymerization and thermodynamic stability of these materials and their recovered new phases by in-situ
high-pressure Raman and $x$-ray diffraction studies. The phonon spectra show that the fullerene molecular
cage in the high-pressure phases is preserved, while these polymers decompose under heat treatment into the
initial fullerene $C_{60}$ monomer. In Ref.~\cite{wang05}  the orthorhombic   polymer have been studied by   
NMR method. Authors conjectured that there exist nine inequivalent carbons on a $C_{60}$ molecule in the orthorhombic  
polymer.\\
It was shown recently in Ref.~\cite{morita11} by the methods  synthetic organic spin chemistry for structurally
well-defined open-shell graphene fragments that graphene, a two-dimensional layer of $(sp^2)$-hybridized 
carbon atoms, can be viewed as a sheet of benzene rings fused together.
Extension of this concept leads to an entire family of phenalenyl derivatives - 'open-shell graphene fragments'.\\
M. Koshino and E  McCann~\cite{kos13} studied the electronic structure of multilayer graphenes 
with a mixture of Bernal and rhombohedral stacking and proposed a general scheme to 
understand the electronic band structure of an arbitrary configuration. The system can be viewed as a series of 
finite Bernal graphite sections connected by stacking faults. They found that the low-energy eigenstates are mostly 
localized in each Bernal section, and, thus, the whole spectrum is 
well approximated by a collection of the spectra of independent sections. The energy spectrum was categorized into 
linear, quadratic, and cubic bands corresponding to specific eigenstates of Bernal sections. The ensemble-averaged 
spectrum exhibits a number of characteristic discrete structures 
originating from finite Bernal sections or their combinations likely to appear in a random configuration. 
In the low-energy region, in particular, the spectrum is dominated by frequently appearing linear bands and quadratic 
bands with special band velocities or curvatures. In the higher-energy region, band edges frequently appear 
at some particular energies, giving optical absorption edges at the corresponding characteristic photon frequencies.\\
Thus molecule-based materials underlie promising next generation
nanoscale electronic devices, machines and
quantum information processing systems, by virtue of
the wide diversity and flexibility in the design of molecular and
electronic structures that can be attained by chemical syntheses.
For example, the spins of unpaired electrons in tailor-made open shell
species can afford control of quantum information in molecules
and thus provide the potential for molecular electronics 
and information processing. \\
Carbon-based materials  and nanostructured materials have huge number of applications~\cite{rod06,emil06,handnano10,hnano10,osawa02,uvar03}. 
The  numerous applications  were presented in the books~\cite{osawa02,uvar03,nanows09,shaf10}, 
which   provides also further information on bioceramics specifically for medical applications~\cite{uvar03,nanows09,shaf10}. New materials for sensors were 
also reviewed. Nanostructured materials and coatings 
for biomedical and sensor applications contribute to the dissemination of state-of-the-art knowledge about the application 
of nanostructured materials and coatings in biotechnology and medicine, as well as that of sensors for the chemical and biomedical industries. The research 
presented in the books~\cite{uvar03,nanows09,shaf10} addresses the fundamental scientific problems that must be resolved in order to take advantage of 
the nanoscale approach to creating new materials.\\
There are various applications of nanoscale carbon-based materials in heavy metal sensing and detection~\cite{analyst}.
These materials, including single-walled carbon nanotubes, multi-walled carbon nanotubes and carbon nanofibers among others, 
have unique and tunable properties enabling applications in various fields spanning from health, electronics and the 
environment sector. Specifically, there are the unique properties of these materials that 
enable their applications in the sorption and preconcentration of heavy metals ions prior to detection by spectroscopic, 
chromatographic and electrochemical techniques. Their unique distinct properties   enable them to be used as novel 
electrode materials in sensing and detection. The fabrication and modification of these electrodes is very fine skill.
Their applications in various electrochemical techniques such as voltammetric stripping analysis, potentiometric stripping 
analysis, field effect transistor-based devices and electrical impedance are numerous.
%
%
%
%
%
%
\section{Carbon-Based Structures and Magnetism}
%
%
%
%
Magnetic properties of carbone-based nanostructures
have attracted a lot of recent interest. 
The permanent magnetic properties of materials such as iron, nickel, cobalt, gadolinium stem from
an intrinsic mechanism of the quantum origin called ferromagnetism. Conventional wisdom has it that
carbon (containing only $s$ and $p$ electrons) does not have a spontaneous magnetic moment in any of its allotropes.
The possibility of ferromagnetism at room temperature in carbon-based materials as, e.g.,
doped graphite, synthetic fullerene $C_{60}$, graphene and carbon composites 
has recently gained a lot of attention of the experimentalists and 
theoreticians~\cite{yazy10,he12,andri03,bouk04,chan04,rode04,talap05,chen08,kan08,pisa08,yang09,kai09,wu09,kroll09,crao09,cer09,yan10,fuku11,bouk11,rao12,gao12}.
The understanding and control of the magnetic properties of carbon-based materials is of fundamental
relevance in their possible applications in nano- and biosciences.\\
Graphite~\cite{bands58,tyler53} has been known as a typical diamagnetic material~\cite{dorf61}. As such it  can be levitated in the strong magnetic 
field. In addition magnetically levitating graphite can be moved with laser and a laser moves the disk in the direction of the light beam. This effect was demonstrated by Kobayashi and Abe~\cite{abe12}. They
showed that the magnetically levitating pyrolytic graphite can be moved in the arbitrary place by simple photoirradiation. 
It is notable that the optical motion control system described in their paper requires only $NdFeB$  permanent magnets 
and light source. 
The optical movement was driven by photothermally induced changes in the magnetic susceptibility of the graphite. Moreover, 
they demonstrated that light energy can be converted into rotational kinetic energy by means of the photothermal property. 
They found that the levitating graphite disk rotates at over 200 $rpm$  under the sunlight, making it possible to develop a new class of light 
energy conversion system.\\
A physical property of particular interest regarding all the
aforementioned carbon allotropes is the magnetic 
susceptibility~\cite{pinni54,suscep55,dsuscep61,hadd91,hadd94,haddon94,msuscep94,msuscep94e,he11,kos11,ando11,kos12},
$\chi_{b}$, since this bulk probe is related to the low energy
electronic spectrum. 
The relationship between the magnetization
induced in a material $\textbf{M}$ and the external field $\textbf{H}$ is defined as:
\begin{equation}
\label{5.1} 
\textbf{M} =  \chi_{b} \textbf{H}.
\end{equation}
The parameter $\chi_{b}$
is treated as the bulk magnetic susceptibility of the material. It can be a complicated
function of orientation, temperature, state of stress, time scale of observation and applied field, but is
often treated as a scalar. 
It is of use  to consider the symbol $\textbf{M}$ for volume normalization
(units of $A m^{-1}$).   Volume normalized magnetization therefore has the same units as $\textbf{H}$.
Because $M$ and $H$ have the same units, $\chi_{b}$ will be dimensionless.\\
Graphite is known as one of the strongest diamagnetic
materials among natural substances.  This property
is due to the large orbital diamagnetism related to
the small effective mass in the band structure, i.e., narrow
energy gap between conduction and valence bands.
The diamagnetic effect becomes even greater in graphene
monolayer  which is truly a zero-gap system. 
R.C. Haddon~\cite{carb2} showed that
the difference in magnetic susceptibility of graphite and diamond prompted Raman to postulate the flow of currents
around the ring system of graphite in response to an applied magnetic field. The discovery of new carbon allotropes,
the fullerenes, has furthered our understanding of this phenomenon and its relationship to aromatic character. $C_{60}$
and the other fullerenes exhibit both diamagnetic and paramagnetic ring currents, which exert subtle effects on the
magnetic properties of these molecules and provide evidence for the existence of $\pi$-electrons mobile in three
dimensions.\\
In general, all known carbon allotropes
exhibit diamagnetic susceptibility~\cite{pinni54,suscep55,dsuscep61,hadd91,hadd94,haddon94,msuscep94,msuscep94e,he11,kos11,ando11,kos12}   
with a few  exceptions. The polymerized
$C_{60}$ prepared in a two-dimensional rhombohedral
phase (depending on the orientation of the magnetic field relative to the polymerized
planes)   shows  weak ferromagnetic signal in some experiments.  Also the disordered
glass-like magnetism  was observed in activated carbon fibers
possibly due to non-bonding $p$ electrons located at edge states. 
The unusual magnetic behavior was observed as well in single
wall carbon nanohorns  which was ascribed to the Van Vleck paramagnetic
contribution.   
Superparamagnetic and/or ferromagnetic-like behavior in
carbon-based material has been previously observed in amorphous
carbon materials obtained by chemical synthesis or
pyrolysis.  The origin of ferromagnetism was suggested
to be attributed to the mixture of carbon atoms with $sp^{2}$ and
$sp^{3}$ bonds and resulted ferromagnetic interaction of spins
separated by $sp^{2}$ centers.  An increase in saturated magnetization
of amorphous-like carbon prepared from different
hydrogen-rich materials indicated the importance of hydrogen
in the formation of the magnetic ordering in
graphite. The ferro- or ferrimagnetic ordering
was reported in proton-irradiated spots in highly oriented pyrolytic graphite. It was demonstrated that protons
implanted in highly oriented pyrolytic graphite triggered ferro- (or, ferri-) magnetic
ordering with a Curie temperature above room temperature.\\
Previously, the "hybrid materials" known as molecular ferromagnets~\cite{enoki,veci01,blund04} in which organic groups are combined with transition metal 
ions were prepared~\cite{veci01,blund04}. Here the organic groups were themselves not magnetic but were used to mediate the 
magnetism between transition metal ions. Organic ferromagnetism was first achieved using organic radicals called nitronyl nitroxides. Many organic radicals exist which have 
unpaired spins, but few are chemically stable enough to assemble into crystalline structures.
Ferromagnetism in organic materials is rare because their
atomic structure is fundamentally different from metals. One of
the few examples identified to date is called TDAE-$C_{60}$: a
compound comprising spherical carbon cages attached to an
organic molecule known as tetrakis-dimethylamino-ethylene.
Since its identification in 1991~\cite{organ91}, many theoretical and
experimental studies have provided some insight into the
mechanism driving this unexpected ferromagnetism~\cite{organ99}, but the
explanation was not fully definitive.\\
It is also possible to prepare molecule based magnets
in which transition metal ions are used to provide the magnetic moment,
but organic groups mediate the interactions.  
In 1998 the ferromagnetism in a cobaltocene-doped fullerene derivative
was reported~\cite{miha98}.  This strategy led to the  fabrication of
magnetic materials with a large variety of structures, including chains, layered
systems and three-dimensional networks, some of which show ordering at
room temperature and some of which have very high coercivity.
Nevertheless, it was recognized that "reports of weak magnetization in organic materials have often proved to be wrong".\\
Magnetism in carbon allotropes has indeed been a fundamental and also controversial problem for a 
long time~\cite{carb1,carb2}. It is of importance to examine this complicated problem thoroughly.\\
In spite of the fact that carbon is diamagnetic,
in 2001 an "observation of strong magnetic signals in rhombohedral pristine $C_{60}$, indicating a Curie
temperature $T_{C}$  near 400-500 K" was reported~\cite{mak1,esq1}. 
In short, it was speculated that the
polymerization of $C_{60}$ fullerenes at certain pressure and temperature conditions, 
as well as photopolymerization in the presence 
of oxygen may lead  to appearance of magnetically ordered phases. "Ferromagnetic behavior" was reported which is 
close to the conditions where the fullerene cages are about to be destroyed, and the effect was presumably 
associated with the defects in intramolecular or intermolecular bonding. In the authors' opinion
the observation of magnetic domain structure in impurity-free regions provides an evidence in favor 
of the \emph{intrinsic nature} of fullerene ferromagnetism.\\
As was  shown above, polymerization of fullerenes can be realized through  various methods~\cite{dav96,dav97,dav97a,dav98,dav00,dav04,mel12} including the high-pressure and temperature treatments and through 
irradiation with $UV$ light.  It can also occur through reactions with alkali metals.
The reported measurements~\cite{mak1,esq1} were described  as the "ferromagnetic fullerene"~\cite{wood02,blund02}.
This new magnetic forms of $C_{60}$ have been identified with the state which occur in the
rhombohedral polymer phase. The existence of previously reported
ferromagnetic rhombohedral $C_{60}$ was confirmed. This property has been shown to
occur over a range of preparation temperatures at $9 \, \textrm{GPa}$. The structure was shown
to be crystalline in nature containing whole undamaged buckyballs. Formation
of radicals is most likely due to thermally activated shearing of the bridging bond
resulting in dangling bond formation. With increasing temperatures this process
occurs in great enough numbers to trigger cage collapse and graphitization.
The magnetically strongest sample was formed at 800 K, and has a saturated
magnetization at $10 K.$\\
Moreover, in paper~\cite{nar03} the observation 
of the ferromagnetically ordered state in a material obtained by high-pressure high-temperature treatment of the 
fullerene $C_{60}$ was confirmed. It had a saturation 
magnetization more than four times larger than that reported previously. From their data  the considerably 
higher value of $T_{C} \approx $ 820 K was estimated~\cite{nar03}.\\
The  widely advertized "discovery" of a ferromagnetic form of carbon~\cite{mak1,esq1,wood02}  stimulated   huge stream 
of  the investigations  of the carbon-based materials~\cite{mak1,esq1,wood02,nar03,han03,mak04,esq2,esq05,mak06,rao12}.
However, difficulties to reproduce those results and the unclear role of impurities casted doubts on the existence
of a ferromagnetic form of carbon.\\
And nevertheless, it was claimed that "the existence of carbon-based magnetic material requires a root-and-branch rework of magnetic theory". 
Moreover, "the existing theory for magnetism in elements with only $s$ and $p$ electron orbits (such as carbon)" should be
reconsidered in the light of the fact that there are many 
publications "describing ferromagnetic structures containing either pure carbon or carbon combined with first row elements",
in spite of the fact that "these reports were difficult to reproduce". \\
It is worth mentioning that in the publications~\cite{mak1,nar03,han03} the  characterization of the samples was not
made properly. This fact was recognized by the authors themselves~\cite{khan03,spem03,mak2}.
In paper~\cite{spem03} 
a $C_{60}$ polymer has been characterized for the first time with respect to impurity content and ferromagnetic properties 
by laterally resolved particle induced $x$-ray emission   and magnetic force microscopy   in order to prove the 
existence of intrinsic ferromagnetism in this material. In the sample studied the main ferromagnetic impurity found was 
iron with remarkable concentration. In spite of that fact authors insisted that they were able
"to separate between the intrinsic and extrinsic magnetic regions and to directly prove that intrinsic 
ferromagnetism exists in a $C_{60}$ polymer".\\
In 2004  the band structure calculations of rhombohedral $C_{60}$ performed in the local-spin-density approximation 
were presented~\cite{bouk04}. Rhombohedral $C_{60}(Rh-C_{60})$ is a 
two-dimensional polymer of $C_{60}$ with trigonal topology. No magnetic solution exists for $Rh-C_{60}$ and energy bands 
with different spins were found 
to be identical and not split. The calculated carbon $2p$ partial density of states was compared to carbon $K$-edge $x$-ray 
emission and absorption spectra and showed good agreement. It was concluded that the rhombohedral distortion of $C_{60}$ 
itself cannot induce magnetic ordering in the molecular carbon. 
The result of magnetization measurements performed on the same $Rh-C_{60}$ sample corroborates this conclusion.\\
It is worth noting that in majority publications on the possible ferromagnetism  of carbon-based "magnetic" material
the effects of the low dimensionality~\cite{may09,kuz10} on the possible magnetic ordering were practically ignored.\\
In 2006 a retraction letter~\cite{mak3} has been published.
Some of the authors (two of them decline to sign this retraction) 
recognized that reported high-temperature ferromagnetism in a
polymeric phase of pure carbon that was purportedly free of
ferromagnetic impurities was an artefact. Other measurements
made on the same and similar samples using particle-induced
$x$-ray emission   with a proton micro-beam have indicated
that these had considerable iron content. Also, polymerized $C_{60}$
samples mixed with iron before polymerization had a similar Curie
temperature (500 K) to those they described~\cite{mak1}, owing to the presence of
the compound $Fe_{3}C$ (cementite). In addition, it has since been
shown that the pure rhombohedral $C_{60}$ phase is not ferromagnetic~\cite{bouk04}.
Nevertheless, they concluded that "magnetic order in impurity-free graphitic structures
at room temperature has been demonstrated independently (before
and after publication of ref.~\cite{mak1}). Ferromagnetic properties may yet be
found in polymerized states of $C_{60}$ with different structural defects
and light-element $(H, O, B, N)$ content".\\
In spite of this dramatic development, the search  for magnetic order at room temperature in a system without the usual $3d$ metallic magnetic 
elements continues. It was conjectured that the 
graphite structure with defects and/or hydrogen appears to be one of the most promising candidates to find this phenomenon.\\
The irradiation effects  for the properties of carbon-based materials were found substantial.
Some evidence that proton irradiation on highly oriented pyrolytic
graphite samples may triggers ferro- or ferrimagnetism was reported~\cite{esq03,esq07}. The possibility of a
magnetism in graphene nanoislands was speculated and a defective graphene phase predicted to be
a room temperature ferromagnetic semiconductor was conjectured as well~\cite{pisa08}.\\
O. V. Yazyev  and L. Helm~\cite{oyde07}
studied from first principles the magnetism in graphene induced by single carbon atom defects. For two
types of defects considered in their study, the hydrogen chemisorption defect and the vacancy defect, the
possibility of the itinerant magnetism due to the defect-induced extended states has been concluded.  
The coupling between  the magnetic moments is either ferromagnetic or antiferromagnetic,
depending on whether the defects correspond to the same or to different hexagonal sublattices of the
graphene lattice, respectively. The relevance of itinerant magnetism in graphene to the high-$T_{c}$ magnetic
ordering was discussed.\\
In Ref.~\cite{rode09}
vacancies and vacancy clusters produced by carbon ion implantation in highly oriented
pyrolytic graphite, and their annealing behavior associated with the ferromagnetism of
the implanted sample were studied using positron annihilation in conjunction with ferromagnetic
moment measurements using a superconducting quantum interferometer device
magnetometer. Authors' results give some indication that the "magnetic moments" may be correlated to
the existence of the vacancy defects in the samples and this is supported by theoretical calculations
using density functional theory. The possible mechanism of magnetic order in the implanted sample was discussed.
Authors~\cite{rode09} claimed that "it has become evident \ldots that even pure carbon can show substantial paramagnetism
and even ferromagnetism".\\
In paper~\cite{esq10} recently obtained data
were discussed  using different experimental methods including magnetoresistance measurements that indicate the 
existence of metal-free high-temperature magnetic order in graphite. Intrinsic as well as extrinsic difficulties to trigger magnetic 
order by irradiation of graphite were discussed. 
The introduction of defects in the graphite structure by irradiation may be in principle a  relevant 
method to test any possible magnetic order in carbon since it allows to minimize sample handling and to estimate 
quantitatively the produced defect density in the structure. The main magnetic effects produced by proton irradiation 
have been reproduced in various further studies. $x$-ray magnetic circular dichroism   studies on proton-irradiated 
spots on carbon films confirmed that the magnetic order is correlated 
to the $\pi$-electrons of carbon only, ruling out the existence of magnetic impurity contributions.
The role of defects and vacancies  continues to be  under current intensive study.\\
In paper~\cite{he12}, by means of
near-edge $x$-ray-absorption fine-structure   and bulk magnetization measurements, it was demonstrated
that the origin of ferromagnetism in $^{12}C^+$ ion implanted highly oriented pyrolytic graphite   is closely
correlated with the defect electronic states near the Fermi level. The angle-dependent 
near-edge $x$-ray-absorption fine-structure spectra imply
that these defect-induced electronic states are extended on the graphite basal plane. It was concluded that  the origin
of electronic states to the vacancy defects created under  $^{12}C^+$  ion implantation. The intensity of the observed
ferromagnetism in highly oriented pyrolytic graphite is sensitive to the defect density, and the narrow implantation dosage window that
produces ferromagnetism should be  optimized.\\
In paper~\cite{ugeda12},
electronic and structural characterization of divacancies in irradiated graphene was investigated.
Authors provided a thorough study of a carbon divacancy, a point defect expected to have a large impact on
the properties of graphene. Low-temperature scanning tunneling microscopy imaging of irradiated graphene
on different substrates enabled them to identify a common twofold symmetry point defect. Authors performed  
first-principles
calculations and found that the structure of this type of defect accommodates two adjacent missing atoms in a
rearranged atomic network formed by two pentagons and one octagon, with no dangling bonds. Scanning
tunneling spectroscopy measurements on divacancies generated in nearly ideal graphene showed an electronic
spectrum dominated by an empty-states resonance, which was ascribed to a nearly flat, spin-degenerated band of
$\pi$-electron nature. While the calculated electronic structure rules out the formation of a magnetic moment around
the divacancy, the generation of an electronic resonance near the Fermi level reveals divacancies as key point
defects for tuning electron transport properties in graphene systems. 
Thus the situation is still controversial~\cite{esq10}.\\
High-temperature ferromagnetism in graphene and other graphite-derived materials reported by 
several workers~\cite{wang09,rao12} 
has attracted considerable interest. Magnetism in graphene and graphene nanoribbons is ascribed to defects and edge states, 
the latter being an essential feature of these materials~\cite{yazy10,guin05,oyde07,ugeda12,mkatz12}. Room-temperature ferromagnetism in graphene~\cite{rao12}  is 
affected by the adsorption of molecules, especially hydrogen. Inorganic graphene analogues formed by some
layered materials  also show such ferromagnetic behavior~\cite{rao12}. Magnetoresistance observed in graphene and 
graphene nanoribbons is of significance because of the potential applications.\\
The problem  of possible intrinsic magnetism of graphene-based materials  was clarified in paper~\cite{sep10}.
The authors have studied magnetization of graphene nanocrystals obtained by sonic exfoliation of graphite. No
ferromagnetism was detected at any temperature down to 2 K. Neither do they found strong paramagnetism
expected due to the massive amount of edge defects. Rather, graphene is strongly diamagnetic, similar to
graphite. Their nanocrystals exhibited only a weak paramagnetic contribution noticeable below 50 K. The
measurements yield a single species of defects responsible for the paramagnetism, with approximately
one magnetic moment per typical graphene crystallite.\\
It should be noted once again that finding the way to make graphite magnetic could be the first step to utilising it as a bio-compatible
magnet for use in medicine and biology as effective biosensors. 
Thus the researchers~\cite{aban11} found a new way to interconnect spin and charge by applying a relatively weak magnetic 
field to graphene and found that this causes a flow of spins in the direction perpendicular to electric current, 
making a graphene sheet magnetized.
The effect resembles the one caused by spin-orbit interaction but is larger and can be tuned by varying the external 
magnetic field. They also show that graphene placed on boron nitride is an ideal material for spintronics because 
the induced magnetism extends over macroscopic distances from the current path without decay. \\
Recently the Geim's team  investigations shed an additional light on controversial magnetic behavior 
of carbon-based structures.~\cite{nair11,sepion12,spema12,sepi12}. 
It was shown~\cite{nair11,sepion12,spema12,sepi12}  that magnetism in many commercially
available graphite crystals should be attributed to micron-sized clusters of predominantly iron.  Those clusters would
usually be difficult to find unless the right instruments were used in a particular way.\\
To arrive at their conclusions,  a piece of commercially-available
graphite was divided  into four sections and  the magnetization of each piece was measured. They found
significant variations in the magnetism of each sample. Thus it was concluded  
that the magnetic response had to be caused by external factors, such as small amount of impurities of
another material  (the induced magnetism). To confirm that essential fact the structure of the samples was thoroughly investigated using a
scanning electron microscope.   It was found that there were unusually heavy particles
positioned deep under the surface.
The majority of these particles were confirmed to be iron and titanium, using a technique known
as $x$-ray microanalysis. As oxygen was also present, the particles were likely to be either
magnetite or titano-magnetite, both of which are magnetic.\\
The very ingenious craftsmanship was used to deduce how many magnetic particles would be needed, and how
far apart they would need to be spaced in order to create the originally observed magnetism~\cite{nair11,sepion12,sepi12}. 
The observations from their experiments agreed with their estimations, meaning the visualized
magnetic particles could account for the whole magnetic signal in the sample.
%
%
%
%
\section{Magnetic Properties of Graphene-Based Nanostructures}
%
%
%
There has been increasing evidence that
localized defect states (or surface or edge states) in $sp$
materials may form local moments and exhibit collective
magnetism ~\cite{yazy10,he12,yang09,kai09,kroll09,cer09,guin05,oyde07,ugeda12,mkatz12}. 
A special interest was connected with the magnetic properties of graphene-based 
nanostructures~\cite{palac07,palac10,chandra11,wang12,hariga01}. Furthermore graphene is a promising candidate
for graphene-based electronics.\\
Ferromagnetism in various carbon structures,
mostly highly defective, has been observed   and investigated theoretically.
Edges of nano-structured graphene, or vacancies, cracks, \emph{etc.}, may lead to localized 
states that increase density of states close to Dirac point.\\
First principles and mean-field theory calculations have shown
that zero-dimensional graphene nanodots or nanoflakes, graphitic petal arrays,
one-dimensional nanoribbons, nano and two-dimensional
nanoholes  that consist of zigzag edges can all exhibit magnetism,
making them an interesting new class of nanomagnets.
It was speculated that the magnetization in graphene-based nanostructures may be originated from the localized
edge states  that give rise to a high density of states at
the Fermi level rendering a spin-polarization instability. 
However, the complete mechanism leading to the magnetic properties  is still a matter of discussion.
Despite the promise shown by the theoretical studies in
magnetic graphene-based nanostructures, however, the experimental realization of
these magnetic graphene-based nanostructures remains a big challenge because synthesis
of graphene is  a difficult task before one further
makes them into different forms of nanostructures~\cite{chen08}.\\
Electronic and magnetic structure of graphene nanoribbons 
and semi-infinite ribbons made from graphene sheets is of especial 
interest~\cite{palac07,palac10,chandra11,wang12,hariga01,tour09,tour11,raotour11,raotour12,tour12}.
It was shown, that the electronic
structure of these ribbons is very sensitive to the edge geometry and to the width
of the ribbon. The strong influence of the exact edge geometry is a
typical feature for graphene, and will also be reflected in the conductance properties of
the nanoribbon~\cite{palac07,palac10,chandra11,wang12,hariga01,tour09,tour11,raotour11,raotour12,tour12}.\\
The review paper~\cite{palac10} covers some of the basic theoretical aspects of the electronic and magnetic
structure of graphene nanoribbons, starting from the simplest tight-binding models to the more
sophisticated ones where the electron-electron interactions were considered at various levels of
approximation. Nanoribbons can be classified into two basic categories, armchair and zigzag,
according to their edge termination, which determines profoundly their electronic structure.
Magnetism, as a result of the interactions, appears in perfect zigzag ribbons as well as in
armchair ribbons with vacancies and defects of different types.
Therefore, the effects of different edges on the
transport properties of nanometer-sized graphene devices need to be investigated carefully. Especially,
it is known, that the two basic edge shapes, namely zigzag and armchair, lead
to different electronic spectra for graphene nanoribbons.\\
K. Harigaya~\cite{hariga01} analyzed theoretically
the mechanism of magnetism in stacked nanographite.
Nanographite systems, where graphene sheets of dimensions of the order of
nanometres are stacked, show novel magnetic properties, such as spin-glasslike
behaviors and change of electron spin-resonance linewidths in the course
of gas adsorptions. Harigaya investigate stacking effects in zigzag nanographite
sheets theoretically, by using a tight-binding model with Hubbard-like on-site
interactions. He found a remarkable difference in magnetic properties between
the simple A-A-type and A-B-type stackings. For the simple stacking, there
were no magnetic solutions. For the A-B stacking, he found antiferromagnetic
solutions for strong on-site repulsions. The local magnetic moments tend to
exist at the edge sites in each layer due to the large amplitudes of the wavefunctions
at these sites. Relations with experiments were discussed.\\
In Ref.~\cite{chen08} using first-principles calculations, it was shown that nanopatterned graphite films   can exhibit
magnetism in analogy to graphene-based nanostructures.   In particular, graphite films with
patterned nanoscale triangular holes and channels with zigzag edges all have ferromagnetic ground
states. The magnetic moments are localized at the edges with a behavior similar to that of graphene-based nanostructures.
Authors' findings suggest that the nanopatterned graphite films form a unique class of magnetic materials.
In conclusion, they have demonstrated that graphite films
can become an all-carbon intrinsic magnetic material when
nanopatterned with zigzag edges, using first-principles calculations.
The magnetism in nanopatterned graphite films may be localized within
one patterned layer or extended throughout all the patterned
layers. It is originated from the highly localized edge states
in analogy to that in graphene-based nanostructures. Because graphite film is readily
available while mass production of graphene remains difficult,
it was argued that the nanopatterned graphite films can be superior for many
applications that have been proposed for graphene-based nanostructures.\\
In Ref.~\cite{fuku11}
the so-called zigzag edge of graphenes has been studied.  It was supposed theoretically  that they has localized 
electrons due to the presence of flat energy bands near the Fermi level. The conjecture was that the localized electron 
spins are strongly polarized, resulting
in ferromagnetism. The  graphenes with honeycomb-like arrays of hydrogen-terminated
and low-defect hexagonal nanopores were fabricated  by a nonlithographic method using nanoporous alumina
templates. Authors  reported large-magnitude room-temperature ferromagnetism caused by electron spins
localizing at the zigzag nanopore edges. This observation may be  a realization of rare-element free,
controllable, transparent, flexible, and mono-atomic layer magnets and novel spintronic devices.\\
In summary, edge atomic structures of graphene
have been of great interest~\cite{palac07,palac10,chandra11,wang12,hariga01,tour09,tour11,raotour11,raotour12,tour12}, due to its strongly
localized electrons, which originate from the presence of flat
energy bands near the Fermi level.
The zigzag edge of graphene   may have  a high electronic
density of states.   The localized edge
electron spins become stabilized and strongly polarized  leading to possible
"ferromagnetism" depending on the exchange interaction
between the two edges, which forms a maximum spin ordering
in these orbitals similar to the case of Hund's rule for
atoms, e.g., the localized edge spins in a graphene nanoribbon
and in graphene with hexagonal nanopore arrays. Moreover, spin
ordering strongly depends on the termination of edge dangling
bonds by foreign atoms (e.g., hydrogen (H)) and those
numbers that result in the formation of edge $\pi$ and $\sigma$
orbitals.\\ 
From another theoretical viewpoint, Lieb's theorem~\cite{lieb89} for bipartite lattices 
predicts that an increase in the difference between the number
of removed A and B sites of the graphene bipartite lattice at
zigzag edges may induce  net magnetic moments and yields ferromagnetism,
particularly in nano-size graphene flakes and nanopores.\\  
It should be stressed that rigorous proof of the appearance of ferromagnetism in realistic model of itinerant electrons
is extremely complicated problem~\cite{kuz09,kuz10,kuz02,tizu72,suto91}.
Lieb's theorem~\cite{lieb89}  
regards the total spin $S$ of the exact ground state
of the  attractive and repulsive Hubbard model  in bipartite lattices. 
Lieb~\cite{lieb89} obtains that independently of lattice structure, for $N$ even the ground state
is unique (and hence has $S = 0$) if $U < 0.$  For $U > 0$ at half-filling on bipartite lattice
the ground state is a $(2S + 1)-$fold degenerate state with $S = 1/2 ||B| - |A||$.
From the other hand, Rudin and Mattis~\cite{rudin85}
calculated the ground state of the Hubbard model in two dimension approximately. They found ground state energies of paramagnetic
and ferromagnetic states as well as the (Pauli) paramagnetic spin susceptibility. Their conclusion was that the 
conditions for ferromagnetism fail to be met in their (non-rigorous) approach.
Moreover,  the rigorous proof of the applicability of the Hubbard model  to  zigzag edges is still lacking as well. 
Thus the real reason of the strong  polarization of electrons was not established definitely.  
The problem of magnetism due to edge states requires  the additional careful investigations and separate thorough discussion.
%
%
%
\section{Some Related Materials}
%
The ferromagnetism is a macroscopic phenomena~\cite{tizu72}. On the microscopic level it is a cooperative effect
of the  Coulomb interaction in many-electron systems and is a consequence of the Pauli exclusion principle.
The term "ferromagnetism" has been used often in   literature in too broad sense.
According to Arrott,  by definition, a material is ferromagnetic if there can exist regions within the
material where a spontaneous magnetization exists. The temperature below which ferromagnetism occurs is called the
Curie temperature and is a measure of the interaction energy associated with the ferromagnetism.
As Arrott~\cite{arrot57}  showed there is \emph{a criterion} for onset of ferromagnetism in a material as
its temperature is lowering from a region in which the linearity  of its magnetic moment versus field isotherm
gives an indication of paramagnetism.\\
While magnetic order is most common in metallic materials
containing narrow bands of $d$- or $f$-electrons, the
magnetic polarization of $p$-electrons has been investigated
only during recent years. Triggered by the growing interest in spintronics materials the search for magnetic semiconductors
or for half-metals used for spin-injection has produced quite a number of new materials classes.\\
The future of the spintronic technology requires the development of magnetic semiconductor
materials~\cite{ques07}. The search for magnetic semiconductors has gathered the
attention of researches for many years. They are rare
in nature, and when they do occur, they possess low Curie temperatures, and their potential technological applications
depend on their capacity to keep the ferromagnetic order up to room temperature.  Most research groups have focused on diluted magnetic semiconductors because of the promising
theoretical predictions and initial results. In the work~\cite{ques07}, the current experimental situation of $ZnO$ based
diluted magnetic semiconductors was considered. Recent results on unexpected ferromagnetic-like behaviour in
different nanostructures were also revised, focusing on the magnetic properties of $Au$ and $ZnO$ nanoparticles
capped with organic molecules. These experimental observations of magnetism in nanostructures without
the typical magnetic atoms are of great importance and were discussed thoroughly. The doubts around the intrinsic origin of ferromagnetism in
diluted magnetic semiconductors along with the surprising magnetic properties in absence of the typical
magnetic atoms of certain nanostructures should make us consider new approaches in the quest for room
temperature magnetic semiconductors.\\  
Magnetism in systems that do not contain transition metal or rare earth ions was recently attracted exceptionally
big attention. There are numerous both the experimental and theoretical results in literature which 
treats the term  "ferromagnetism" too broadly and without the proper carefulness.
Magnetism is a cooperative phenomenon essentially and can be an \emph{intrinsic} property of a crystalline state, 
or it can be \emph{induced} by magnetic impurities in a non-magnetic system or by  specific defects or vacancies.
The main problem in this case is the fact that spin polarization is local.  The possibility of the magnetically ordered
state will depend on the delicate balance of the various interactions in the system.
The non-vanishing stable spin polarization may arise in   certain cases due to specific combination of peculiarities
of the band structure of the system but as an exception. We summarize  very briefly a few examples below.\\
Okada and Oshiyama~\cite{okad01} reported first-principles total-energy electronic-structure calculations in the
density functional theory performed for hexagonally bonded honeycomb sheets consisting of $B$, $N$, and  $C$ atoms.
They found that the ground state of $BNC$ sheets with particular stoichiometry is ferromagnetic. Additional analysis
of energy bands and spin densities leads them to conclusion that the nature of the ferromagnetic ordering is
connected with the flat-band character of electronic structure. The flat-band ferromagnetism~\cite{wu09} is one of the of the
mechanisms which may play a role side by side with the high density of states at the Fermi level and strong electron
correlation.\\
In   paper~\cite{dev08} cation-vacancy induced intrinsic magnetism in $GaN$ and $BN$ was investigated by employing density functional
theory based electronic structure methods. It was shown that the strong localization of defect states favors
spontaneous spin polarization and local moment formation. A neutral cation vacancy in $GaN$ or $BN$ leads
to the formation of a net moment of $3\mu_{B}$ with a spin-polarization energy of about 0.5 $eV$ at the low density
limit. The extended tails of defect wave functions, on the other hand, mediate surprisingly long-range
magnetic interactions between the defect-induced moments. This duality of defect states suggests the
existence of defect-induced or mediated collective magnetism in these otherwise nonmagnetic $sp$
systems.\\
The $p$-electron magnetism in doped $BaTiO_{3-x}M_{x}$ ($M = C,N, B$) was investigated in Ref.~\cite{grub12} 
Authors presented Vienna
$ab$ initio simulation package (VASP) calculations using the the hybrid density functional  for carbon, nitrogen, and
boron-doped $BaTiO_{3-x}M_{x}$ ($M = C,N, B$). They calculated a 40-atom supercell and replaced one oxygen
atom by  $C$, $N$, or $B$. For all three substituents they  found a magnetically ordered ground state which
is insulating for $C$ and $N$ and half-metallic for $B$. The changes in the electronic structure between
the undoped and the doped case are dominated by the strong crystal field effects together with the
large band splitting for the impurity $p$-bands. Using an $MO$ picture they proposed an explanation for the
pronounced changes in the electronic structure between the insulating non-magnetic state and
the as well insulating magnetic state for doped $BaTiO_{3}$. The conclusion was made that the $p$-element-doped 
perovskites could provide a new class of materials for various applications ranging from spin-electronics to 
magneto-optics.\\
In   interesting paper by Upkong and Chetty~\cite{ukpong} the  study of substitutionally doped boronitride was 
carried out. 
They performed first-principles molecular dynamics simulations to investigate the magnetoelectronic response of 
substitutionally doped boronitrene to thermal excitation. Authors showed that the local geometry, size, and edge 
termination of the substitutional complexes of boron, carbon, or nitrogen determine the thermodynamic stability of the 
monolayer. 
In addition, they  found that hexagonal boron or triangular carbon clusters induce finite magnetic moments 
with 100$\%$ spin-polarized Fermi-level electrons in boronitrene. In such carbon substitutions, the spontaneous 
magnetic moment increases with the size of the embedded 
carbon cluster, and results in half-metallic ferrimagnetism above $750 \, K$ with a corresponding Curie point of $1250 \, K$, 
above which the magnetization density vanishes. Authors predicted an ultrahigh temperature half-metallic ferromagnetic 
phase in impurity-free boronitrene, when any three nearest-neighbor nitrogen atoms are substituted with boron, 
with unquenched magnetic moment up to its melting point.\\
Unfortunately the form of the presentation of their results contains some delicate misleading features related to the
precise definition of the  term  "ferromagnetism".
The title "half-metallic ferromagnetism" suggests the long-range order, which it is not present really. What 
authors described were the local high-spin defects where a few spins interact by "ferromagentic coupling". 
But there are another possibility which authors ignored. The nearly the 
same result would be obtained when the clusters were terminated by e.g. hydrogens. 
There is only a slight spin polarization which extends over the rim of the defect. 
Such things  can hardly be termed by "ferromagnetism". Moreover, the authors even used  term "ferrimagnetism".
They found spin-polarized density of states (figure 8) for $s$ and $p$ orbitals of the four carbon atoms in 
the star-shaped carbon cluster. However
the interpretation of the changes of bond length in their figure 8 may be simply Jahn-Teller type distortions which 
are common in open-shell systems. In this sense the paper  treats the term  "ferromagnetism" too broadly and without
the proper carefulness. Nevertheless, the paper is rather stimulating and will promote further investigations in this direction. 
%
\section{Conclusions}
%
In summary, in the present work, the problem of the existence of carbon-based magnetic material
was analyzed and reconsidered to elucidate the possible relevant mechanism (if
any) which may be responsible for observed peculiarities of the "magnetic" behavior in these
systems, having in mind the quantum theory of magnetism criteria. Some theoretical conjectures and
experimental results were re-examined critically. It is difficult to take into account and  to summarize concisely 
the many important results covered in the numerous publications. But we hope that the present study  shed some light on many complicated
aspects of the controversial problem of magnetism of the carbon-based structures. However, it leaves many open questions
which should be answered in additional publications.\\
In general, the origin of magnetism lies in the orbital and spin motions of electrons and how the electrons interact 
with one another~\cite{kuz09,kuz10}. The basic object in the magnetism of condensed matter
is the magnetic moment.  The magnetic moment in practice may depend 
on the detailed environment and additional interactions such as spin-orbit, screening effects and crystal fields.
To understand the full connection between magnetism and chemical structure of a material, a detailed characterization 
of the samples studied is vital.\\
Characterization of magnetic materials by
means of neutron scattering technique is highly desirable~\cite{kuz81,yzhu,ml71,tapa06,fitz04}.
Carbon-based structures  should be investigated by neutron scattering and its spin density
distribution should be measured in order to visualize the pathway of the  magnetic interactions. The spin density
of these complex structures should be measured by polarized neutron diffraction.\\
The similar problem was mentioned already in context of organic ferromagnets~\cite{akita99,maram11}.
Unpaired electrons in these compounds are usually valence electrons and are of great
chemical interest. The polarized neutron diffraction   experiments
were  usually analyzed in such a way as to produce the unpaired electron
density or spin density. The spin density is, or should
be, a positive quantity because (i) there is a higher population of up-spin
electrons than down-spin electrons (when an magnetic field is
applied), and (ii) electrons tend to always be paired so that only the
unpaired electrons contribute to the spin density (i.e. the spin density
from the paired electrons cancels out).\\
Unfortunately the polarized neutron diffraction experiment does not directly determine
the unpaired electron density. Instead, polarized neutrons are scattered
from the magnetic field density in the crystal (they are also scattered
from the nuclei, but this effect can usually be modelled). Therefore,
one should examine the magnetic field density rather than the spin
density itself~\cite{maram11}.   It is of importance to obtain the fully information
on the actual spin density in carbon-based materials. The full picture of the spin density distribution
will clarify the problem of magnetism of the carbon-based materials greatly.\\
On the basis of the present analysis the
conclusion can be  made that the thorough and detailed experimental studies of this problem only may
lead us to a better understanding of the very complicated problem of magnetism of carbon-based
materials.
%
%
%
%
\section{Acknowledgements}
%
The author acknowledges A.V. Rakhmanina,  E. Roduner, E. Osawa, O. Yaziev and J. Fernandez-Rossier for valuable discussions, 
comments and useful information.
%
%


%

\end{document}